\documentclass{aa}
\usepackage{CJKutf8}
\usepackage{graphicx}
\usepackage{caption}
\usepackage{booktabs}
\usepackage{multirow}
\usepackage[varg]{txfonts}
\usepackage{subcaption}
\usepackage{float}
\usepackage{soul}
\usepackage[normalem]{ulem}
\renewcommand\thesubfigure{(\alph{subfigure})}
\usepackage[colorlinks=true,
            linkcolor=blue,  
            citecolor=blue,  
            urlcolor=blue,
            breaklinks=true]{hyperref}

\title{Nebular\_AGN: A CIGALE module for fitting active galactic nucleus emission lines}
\titlerunning{Modelling AGN emission lines}

\author{Hao Zhang (\begin{CJK}{UTF8}{gbsn}张浩\end{CJK})\inst{1} \and Patrice Theul\'e \inst{1} \and V\'eronique Buat \inst{1} \and Denis Burgarella \inst{1} \and  Estelle Pons \inst{1}  \and M\'ed\'eric Boquien \inst{2} }
\institute{Aix Marseille Univ, CNRS, CNES, LAM, Marseille, France\\
              \email{hao.zhang@lam.fr}       
        \and  Université Côte d’Azur, Observatoire de la Côte d’Azur, CNRS, Laboratoire Lagrange, Nice, France 
           }             
\begin{document}

\abstract {} {The increasing discovery of high-redshift active galactic nuclei (AGNs) in recent years imposes more stringent requirements on spectral analysis tools for deriving the properties of AGNs and their host galaxies from emission-line diagnostics. To address this need, we have developed a new module for the popular spectral energy distribution (SED) fitting tool Code Investigating GALaxy Emission (CIGALE), namely the \texttt{[nebular\_AGN]} module, which enables the efficient and flexible simulation and fitting of emission lines originating from the broad-line regions (BLRs) and narrow-line regions (NLRs) of AGNs and allows for the estimation of the physical properties of these regions.} {We used the spectral synthesis code \texttt{Cloudy} to construct the database for the new module. Based on the X-ray and accretion disk continua implemented in CIGALE, we generated the incident radiation fields of the models. We then adopted the AGN geometry and dust settings implemented in CIGALE to define a flexible set of physical parameters for the gas clouds, thereby producing a comprehensive database for the \texttt{[nebular\_AGN]} module.} {We benchmarked the \texttt{[nebular\_AGN]} module using a quasar composite spectrum, an empirical metallicity calibration, and observational data from X-ray-selected AGNs. Our module can approximately reproduce the majority of quasar emission-line features, cover the key emission-line ratios observed in AGN samples, and provide an assessment of their physical properties. For specific combinations of parameters, the metallicity derived by our module is consistent with the empirical formula. We further compared our models with other photoionization models used to simulate AGN NLR emission, and we performed a line-sensitivity study to identify the most effective diagnostic lines for each parameter in our module. Finally, we find that the dust attenuation law plays an important role in the SED fitting.}{}


\maketitle
\nolinenumbers
\section{Introduction}
\label{sec introduction}
Active galactic nuclei (AGNs) have long been central to our understanding of galaxy evolution and black hole growth. As the most efficient energy sources in the Universe, supermassive black holes (SMBHs) convert the rest-mass energy of accreted material into powerful radiation covering the entire electromagnetic spectrum, making them some of the brightest astronomical objects. The structure of AGNs and the origins of their emission across different wavelength bands have been extensively studied \citep{Antonucci93, Urry95, NetzerARAA15, Padovani2017}. The accretion disk produces most of the ultraviolet (UV) and optical radiation, while X-ray emission arises from the corona located very close to the central engine. The dust torus, heated by the radiation from the accretion disk and corona, dominates the infrared output \citep{NetzerARAA15, Honig2019}. Relativistic jets make AGNs luminous in the radio band through synchrotron radiation. Most of the observable AGN emission lines originate from the broad-line region (BLR) and the narrow-line region (NLR), which consist of gas clouds surrounding the central black hole and are photoionized by radiation from both the AGN accretion disk and the corona.

Evidence from observations and simulations indicates that the formation and growth of SMBHs are closely related to the nature and evolution of their host galaxies and that SMBH feedback plays a key role in regulating the star formation histories in their host galaxies \citep{Marconi2003,Magorrian1998,Springel2005,Hopkins2006}. The James Webb Space Telescope (JWST) is extending our view into the early Universe, with an increasing number of high-redshift AGNs and AGN candidates being discovered \citep{Juodzbalis2023,Lambrides2024,Scholtz2025,Napolitano2025,Treiber2025}. JWST has also discovered the little red dots at high redshift, which are thought to play an important role in the birth of massive black holes \citep{Harikane2023, Kocevski2023, MattheeApJ24}. All of these discoveries challenge our existing models of galaxy evolution and place greater demands on our ability to identify and diagnose the properties of AGNs through spectral emission lines, especially at high redshift. 

As next-generation wide-field multi-object spectrographs such as VLT/MOONS \citep{MOONS} and Subaru/PFS \citep{PFS} are coming into operation, the urgent need for efficient and accurate computational tools to identify and diagnose potential AGNs from large spectroscopic datasets becomes evident. Such tools will be essential for advancing our understanding of the population distribution and evolution of high-redshift AGNs.

Many spectral energy distribution (SED) fitting codes have been developed to infer the physical properties of galaxies from photometric and spectroscopic data, among them (to name a few) are LePhare \citep{ArnoutsMNRAS99, IlbertAA06}, HYPERZ \citep{HYPERZAA00}, CIGALE \citep{BurgarellaMNRAS05, NollAA09, BoquienAA19}, EAZY \citep{EAZY}, MAGPHYS \citep{MAGPHYS}, BEAGLE \citep{BEAGLE}, PROSPECTOR \citep{PROSPECTOR}, FAST \citep{FAST}, and BAGPIPES \citep{BAGPIPES}. Some of the codes are based on Bayesian inference and the conservation of the energy budget (i.e. the energy emitted by dust in the mid- and far-IR corresponds to the energy absorbed by dust in the UV-optical range). These tools are capable of rapidly generating physically realistic galaxy spectral models spanning from the far-ultraviolet to the microwave regime and fitting them to both spectroscopic and photometric observations to perform a Bayesian analysis of galaxy physical properties. On the AGN side, AGNfitter \citep{AGNfitter2016, AGNfitter2024} is a Bayesian Markov chain Monte Carlo approach designed to fit the SEDs of AGNs and galaxies. It includes six physical emission components: the X-ray corona, an accretion disk, a torus of AGN heated dust, stellar populations, cold dust in star-forming regions, and synchrotron emission from the AGN and star-forming regions. However, AGNfitter focuses exclusively on fitting the global SED and does not model galaxy or AGN emission lines explicitly. The PROSPECTOR extension includes an empirical AGN emission-line template that describes dust torus emission using two parameters \citep{Johnson2021}, but it does not provide constraints on the physical properties of the BLRs and NLRs. The BEAGLE-AGN code \citep{VidalGarciaMNRAS24} extends BEAGLE by incorporating nebular emission from the NLR, but the contribution of the BLR is not included, which means it can only be applied to type II AGNs. NebulaBayes \citep{Thomas2018a, Thomas2018b} is a Bayesian code that derives physical properties by comparing observed emission-line fluxes with photoionization model grids for NLRs and \ion{H}{ii} regions. However, similar to BEAGLE-AGN, the contribution from the BLR is not included.

The Code Investigating GALaxy Emission (CIGALE) is a state-of-the-art Python code for SED fitting of extragalactic sources, and it has undergone continuous development over the past years \citep{BoquienAA19,YangMNRAS20, Yang2022, TheuleAA24, Burgarella2025}. The current version of CIGALE can fit photometric data from the X-ray to radio bands and extract the physical properties of galaxies by combining multiple emission components, such as the stellar populations and their surrounding \ion{H}{ii} regions, dust attenuation with re-emission, and AGN accretion disks. Although nebular emission lines from gas clouds surrounding the young stellar population have been incorporated by the \texttt{[nebular]} module and the AGN continuum has already been modelled through the \texttt{[AGN]} module, CIGALE does not include emission from the BLR nor the NLR of AGNs. As a consequence, it has so far been unable to directly and accurately simulate and fit emission-line contributions from AGNs, thereby limiting its ability to be used to robustly infer AGN physical properties.

In this work, we present a new module for the CIGALE SED fitting code, the \texttt{[nebular\_AGN]} module, which enables CIGALE to infer physical properties of AGNs based on observed emission lines. We used the \texttt{Cloudy v23.01} photoionization code \citep{CLOUDY} to perform radiative transfer simulations for both the BLR and the NLR over a wide range of physical conditions. Both nebular continua and emission lines were incorporated into the CIGALE spectral database. In Section~\ref{sec: model}, we describe in detail the settings of the model used to simulate nebular emission from the BLR and the NLR. In Section~\ref{sec: benchmark}, we use observational data and the empirical relation to benchmark the performance of the new module. In Section~\ref{sec: discussion}, we compare our models with other photoionization models, discuss the sensitivity of different AGN emission lines to various physical parameters, and assess the impact of dust attenuation on line ratios. We summarise the paper in Section \ref{sec: conclusion}.

\section{The AGN nebular model}
\label{sec: model}

In this section, we describe in detail the model of BLR and NLR we use to build the spectral database for the [\texttt{nebular\_AGN}] module, including the geometry of our AGN model, the incident radiation fields, the chemical abundances of elements, and the dust settings. Based on these settings, we performed radiative transfer simulations using the \texttt{Cloudy v23.01} photoionization code along with the \texttt{PyCloudy} interface \citep{pyCloudy} to construct a spectral database containing both continuum and discrete line emission from the BLRs and NLRs of the AGNs. 

\subsection{Geometry}
\label{sec: model/geometry}
\begin{figure}[h!]
\centering
\includegraphics[width=9cm]{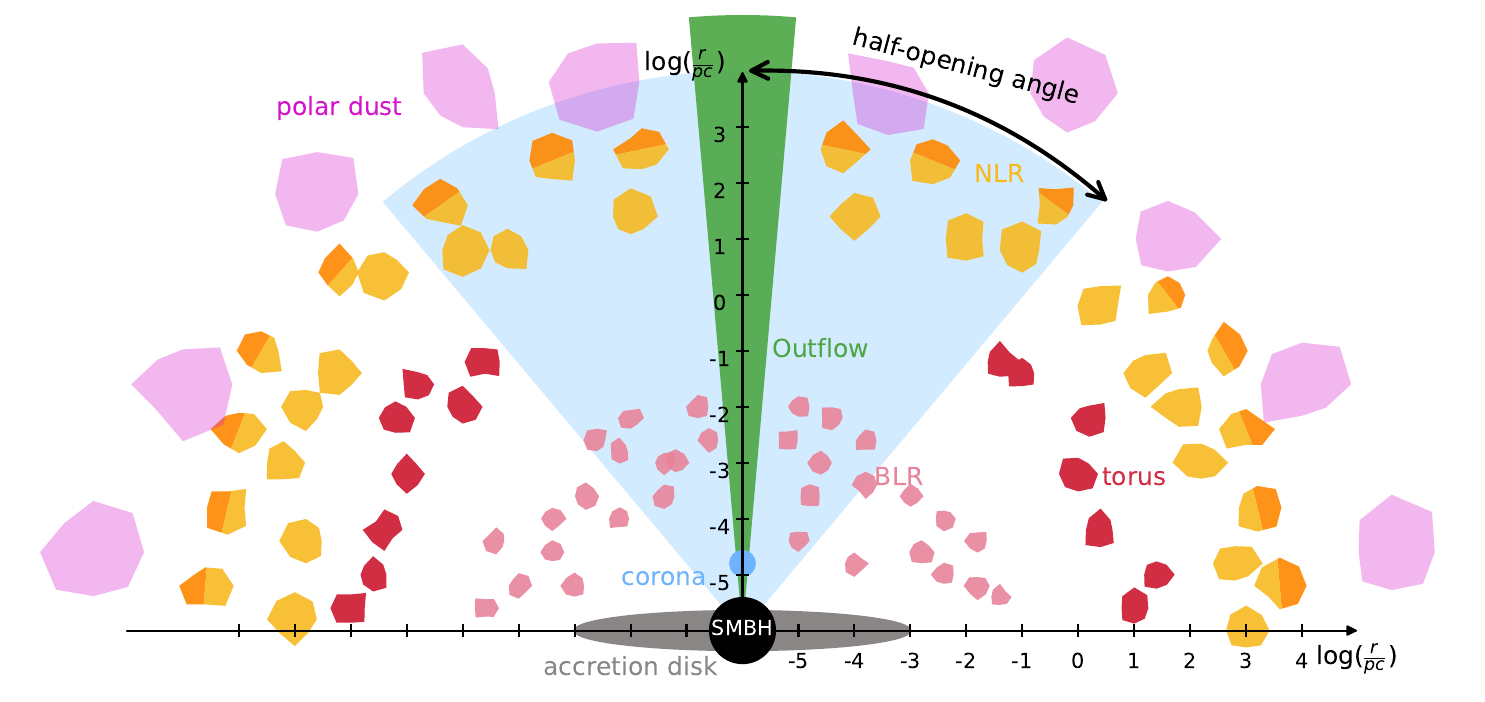}
\caption{Sketch of the AGN geometrical structure used in the [\texttt{nebular\_AGN}] module. The colours are as follows: black: central black hole; grey: accretion disk; red: the dust torus; blue: the corona; pink: BLRs; light orange: the ionized regions in the NLRs; dark orange: PDRs and molecular regions in NLRs (not considered in our model); violet: polar dust; green: outflows (not considered in our model); blue shaded region: opening angle.}
\label{Fig:geometry}
\end{figure}

Figure \ref{Fig:geometry} shows the geometry of the AGN model used in this work. According to the unified model of AGN \citep{Antonucci93, Urry95}, BLRs are clouds orbiting above the disk that are located at distances of approximately 0.001–0.1 pc from the central black hole. The dust torus may obscure the emission from BLRs (type II AGNs) or not (type I AGNs), depending on the relation between the viewing angle (i.e. the inclination between the line of sight and the normal to the accretion disk) and the opening angle (the blue shaded region in Fig.~\ref{Fig:geometry}). Continuum and broad-line emission can be scattered  by hot electrons that pervade the region. The NLRs extend approximately from 100 to 1000 pc \citep{Almeida2017}. However, there is still no consensus on many details of this model. For example, \citet{Nenkova2008} proposed that the dust torus is actually a natural extension of the BLR, and that X-ray absorption, broad-line emission, dust absorption, and infrared re-emission all originate from the same continuously distributed gas cloud system, the so-called toroidal obscuration region (TOR). X-ray and optical obscuration in individual sources are related to the number of dusty clouds along radial equatorial rays, the torus angular thickness, and the optical depth of each cloud. 

In recent years, with the progress of mid-IR interferometry, warm dust distributed along the AGN polar direction has been found to be very common \citep{Asmus2016, Lopez2016}. \citet{Buat2021} discussed the influence of polar dust on the SED of AGN host galaxy and explored the possible shape of its extinction law. For type I AGNs, \citet{YangMNRAS20} used the polar dust to attenuate the disk emission. However, introducing polar dust attenuation for NLR emission is controversial, because the distribution of polar dust is very complex and varies significantly among objects, ranging from scales of a few parsecs \citep{Lyu2018} to several kiloparsecs \citep{Zou2019}. The relationship between polar dust and the spatial distribution of BLRs and NLRs is therefore uncertain, and it is also unclear whether polar dust extends to regions closer to the disk. For type II AGNs, the situation becomes even more complicated because they essentially arise from dust torus obscuration of BLR emission. As a simplifying assumption aimed at limiting the complexity of the model, we assume that polar dust is distributed isotropically around the disk, so polar dust attenuates both the BLR and NLR emission in our model.

The accretion disk, as the primary radiation source, has a geometry consistent with the \texttt{[skirtor2016]} option of the \texttt{[AGN]} module (see Sect.~\ref{sec: model/inciden radiation field/acrretion disk}) in CIGALE which is used to describe the accretion disk continuum. It is approximated as a central point source with anisotropic emission, as described by \citet{Netzer1987}:
\begin{equation}
\mathrm{L(\theta)} \propto \cos\theta(2\cos \theta + 1).
\label{eq:geometryofdisk}
\end{equation}
Here, $\theta$ is the polar angle of the coordinate system, i.e.\ the inclination (or viewing) angle $i$ in Table \ref{Table:parameters}. The disk continuum is further processed by the dust torus. The torus is modelled as a three-dimensional, two-phase medium based on hydrodynamical simulations, with high-density clumps embedded in a low-density medium that fills the space between the clumps \citep{StalevskiMNRAS12, StalevskiMNRAS16} and is implemented in \texttt{[skirtor2016]}.

We adopted the closed geometry in \texttt{Cloudy v23.01}, in which the central object is small relative to the cloud, and all diffuse radiation escaping from the illuminated face of the cloud in the direction towards the central object subsequently interacts with the far side of the cloud \citep{hazy}. This setup simulates scattering between clouds.

We stopped the radiative transfer calculations at the ionization front by imposing a stopping criterion of $\frac{\mathrm{H^{+}}}{\mathrm{H^{0}}} = 0.01$. This implies that the denser regions (the dark orange regions in Fig.~\ref{Fig:geometry}) corresponding to photodissociation regions (PDRs) and molecular regions are not taken into account. As a consequence, the intensities of some low-ionization lines, neutral atomic lines, and molecular emission lines are suppressed to varying degrees, depending on the contribution of the dense regions to the production of these lines.

For the kinematics of the clouds, we introduce two free parameters, \texttt{lines\_width\_NLR} and \texttt{lines\_width\_BLR}, which specify the widths of the Gaussian emission-line profiles in the narrow-line and broad-line regions (in km s$^{-1}$), respectively, under the assumption that all BLRs share the same velocity dispersion and all NLRs share the same velocity dispersion.

In \texttt{[skirtor2016]}, the effects of the dust torus, the opening angle, and the inclination angle on the accretion disk luminosity are already taken into account. Since we use the ionizing photon luminosity $\mathrm{Q(H)}$, calculated based on the accretion disk luminosity provided by \texttt{[skirtor2016]}, to scale the contributions from the BLR and NLR (see Sect.~\ref{sec: model/inciden radiation field/intensity}), the effects of the dust torus, the opening angle, and the inclination angle on the BLR and NLR contributions are therefore implicitly propagated in \texttt{[nebular\_AGN]}. Consequently, no additional treatment is required. In addition, \texttt{[nebular\_AGN]} introduces two free parameters, \texttt{f\_BLR} and \texttt{f\_NLR}, which represent the covering factors of the BLR and the NLR, respectively, and take values between 0 and 1. They are defined as the ratio of the solid angle subtended by the gas, as seen from the central radiation source, to $4\pi$.

\subsection{The incident radiation field}
\label{sec: model/inciden radiation field}
\begin{figure}[h!]
\centering
\includegraphics[width=8cm]{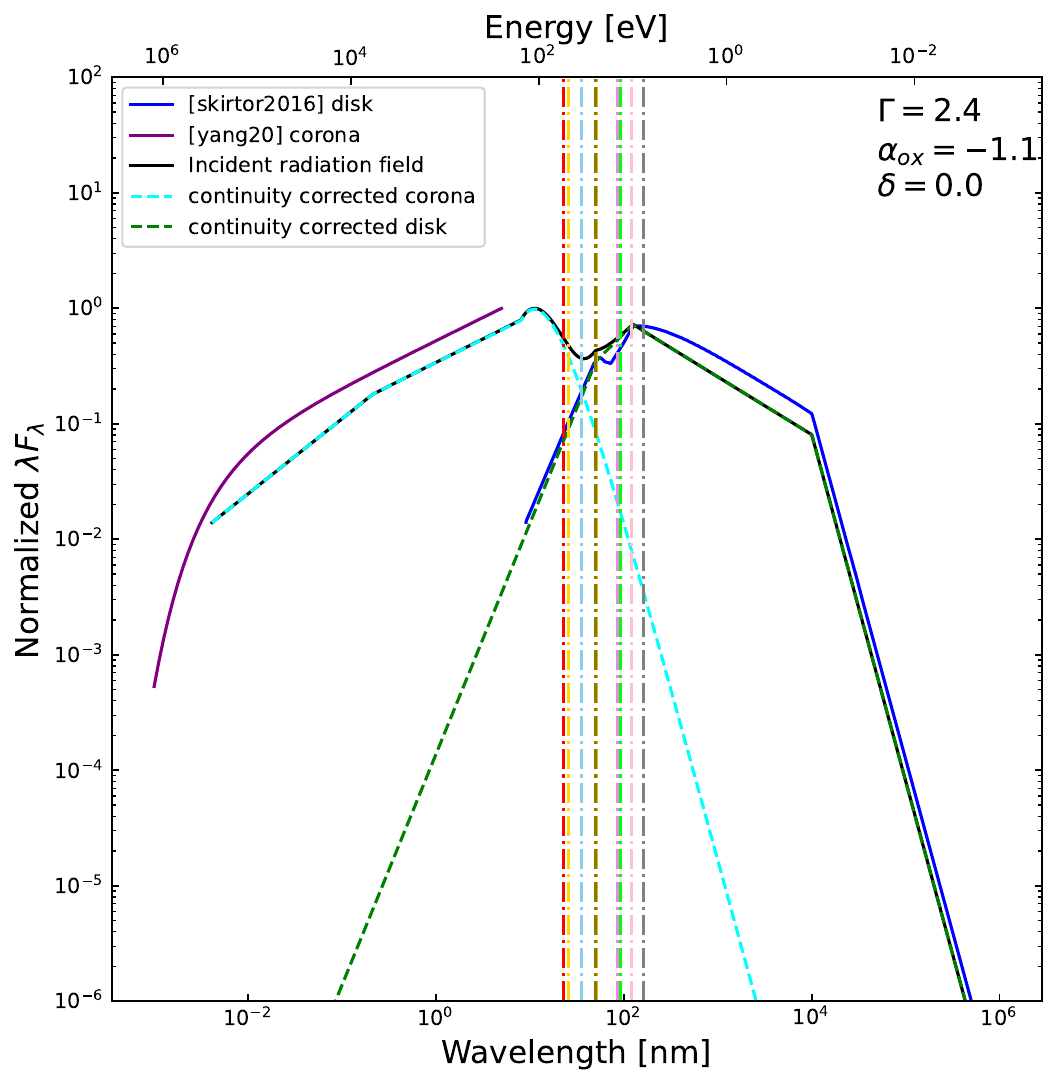}
\caption{Incident radiation field used in the radiative transfer simulations to build the spectral database of the \texttt{[nebular\_AGN]} module. The purple solid line shows the hot coronal emission given by the \texttt{[X-ray]} module (\texttt{[yang20]}, $\mathrm{\Gamma} = 2.4$, $\mathrm{\alpha_{OX}} = -1.1$, $\mathrm{E_{cut}}= 300$~keV). The blue solid line shows the accretion disk emission given by the \texttt{[AGN]} module (\texttt{[skirtor2016]}, Schartmann model, $\delta$ = 0). These two components combined serve as the incident radiation field for BLR and NLR gas clouds. However, we can clearly see a pronounced discontinuity in the continuum independently produced by the two modules over the 5--50 nm range, as discussed in Section~\ref{sec: model/Fixing the discontinuity}. The cyan dashed line and the green dashed line represent the corona continuum and disk continuum after applying the continuity correction, respectively. The solid black line represents the total SED resulting from the sum of the corrected corona and disk continuum, which is the incident radiation field we actually used in photoionization simulations. The vertical dash-dotted lines mark the ionization potentials of a series of atomic and ionic species (H: black; C$^{2+}$: gold; C$^{+}$: orange; He: olive; He$^{+}$: red; Mg: grey; O: lime; O$^{+}$: sky-blue; N: violet; S: pink).}
\label{Fig:incident radiation field}
\end{figure}

The incident radiation field impinging on the BLRs and NLRs in the simulation has both a shape and an intensity. It consists of two main components: the accretion disk emission, which dominates the UV-to-infrared spectrum, and the coronal emission, which dominates the X-ray, as illustrated in Fig.~\ref{Fig:incident radiation field}.

In CIGALE, the X-ray coronal continuum emission is set by the \texttt{[X-ray]} module, while the accretion disk continuum emission is set by the \texttt{[AGN]} module. The shape of the incident radiation field that ionizes both the BLRs and NLRs is specified by these two modules.

\subsubsection{Intensity of the incident radiation field}

\label{sec: model/inciden radiation field/intensity}
The intensity of the incident radiation field impinging on the illuminated face of the gas cloud is quantified by the dimensionless ionization parameter \textbf{U}:
\begin{equation}
\mathrm{U} \equiv\frac{1}{\mathrm{n_H}\mathrm{c}}\int_{\nu_0}^{+\infty}\frac{\mathrm{F_\nu}}{\mathrm{h}\nu} \, \mathrm{d}\nu = \frac{\mathrm{Q(H)}}{4 \pi \mathrm{r^{2}}\mathrm{n_H} \mathrm{c}}=\frac{\mathrm{\Phi(H)}}{\mathrm{n_H} \mathrm{c}},
\label{eq:logU}
\end{equation}
where $\mathrm{F_\nu}$ is the surface energy flux of the incident radiation field, integrated from $\mathrm{h\nu_0}$ = 13.6 eV to infinity, $\mathrm{Q(H)}$ $\mathrm{[s^{-1}]}$ is the intrinsic hydrogen-ionizing photon luminosity emitted by the central excitation source,  $\mathrm{r}$ $\mathrm{[cm]}$ is the distance from the central excitation source to the illuminated face of the cloud, and $\mathrm{\Phi(H)}$ $\mathrm{[cm^{-2} \, s^{-1}]}$ is the surface flux of hydrogen-ionizing photons striking the illuminated face of the cloud. Following the methodology used to build the \texttt{[nebular]} module \citep{BoquienAA19}, we used $\mathrm{\Phi(H)}$ to normalise the simulated continua and line intensities of the BLR and NLR generated by \texttt{Cloudy v23.01} simulation and then rescaled them to the appropriate level by multiplying the intrinsic hydrogen-ionizing photon luminosity $\mathrm{Q(H)}$ of the central AGN. Because of their different densities and distances from the accretion disk, the ionization parameters of the BLR and NLR are expected to differ.

In CIGALE, the value of $\mathrm{Q(H)}$ is constrained by the input observational data to be fitted and parameter settings of the \texttt{[X-ray]} and \texttt{[AGN]} modules. For each model SED generated according to the parameter settings in \texttt{[X-ray]} and \texttt{[AGN]} and involved in the Bayesian-like fitting procedure, CIGALE computes its normalised bolometric luminosity and rescales it according to the input photometry data to obtain the corresponding $\mathrm{Q(H)}$. Since the ionization parameter \textbf{U} and the hydrogen density $\mathrm{n_{H}}$ are also treated as free parameters in the  \texttt{[nebular\_AGN]} module, this effectively implies that, for a given $\mathrm{n_{H}}$, varying \textbf{U} is equivalent to changing the distance between the gas cloud and the AGN disk, which is treated as a point source.

\begin{table}[ht]
\caption{Typiacl distances of the BLR and NLR.}
\centering
\begin{tabular}{lcccc}
\toprule
 region & $\mathrm{r_{-1}}$ & $\mathrm{r_{-2}}$ & $\mathrm{r_{-3}}$ & $\mathrm{r_{-4}}$ \\ & (pc) & (pc) & (pc) & (pc) \\
\midrule

NLR & 48.184 & 152.370 & 481.836 & 1523.698 \\
\cmidrule(lr){1-5}

BLR & 0.015 & 0.048 & 0.152 & 0.482 \\
\cmidrule(lr){1-5}

\end{tabular}
\tablefoot{Typical distances of the BLR and NLR from the central black hole for log\textbf{U} = -1, -2, -3, and -4, assuming $\mathrm{n_{H,NLR}} = 10^{3}\ \mathrm{cm^{-3}}$, $\mathrm{n_{H,BLR}} = 10^{10}\ \mathrm{cm^{-3}}$, and ${\mathrm{L_{bol}}} = 10^{45}\ \mathrm{erg\ s^{-1}}$. Most of the results fall within the expected ranges, with the BLR located at distances of approximately 0.001–0.1 pc from the central black hole, while the NLR extends from about 100 to 1000 pc, as we mentioned in Section~\ref{sec: model/geometry}.}
\label{Table:distance}
\end{table}
In Table~\ref{Table:distance}, we present the distances $\mathrm{r}$ corresponding to log\textbf{U} = -1, -2, -3, and -4, assuming an AGN with bolometric luminosity of $10^{45}\ \mathrm{erg\,s^{-1}}$, $\mathrm{n_{H,NLR}} = 10^{3}$ $\mathrm{cm^{-3}}$, and $\mathrm{n_{H,BLR}} = 10^{10}$ $\mathrm{cm^{-3}}$, based on Eq.~\ref{eq:logU}. Since $\mathrm{Q(H)}$ is proportional to ${\mathrm{L_{bol}}}$, with these reference values, the distances of the BLR and NLR for other choices of $\mathrm{L_{bol}}$ and $\mathrm{n_H}$ can be readily computed using Eq.~\ref{eq:logU} and compared with the typical distances of the BLR and NLR from the central black hole.

\subsubsection{The accretion disk emission}
\label{sec: model/inciden radiation field/acrretion disk}
The \texttt{[AGN]} module of CIGALE, which is used to model the continuum from the accretion disk, provides two options: \texttt{[fritz2006]} and \texttt{[skirtor2016]}. The former is based on the work of \citet{Fritz2006}, who performed radiative transfer calculations taking into account three main components: the primary source located in the torus, the scattered emission by dust, and the thermal dust emission. The latter includes three different models: Schartmann \citep{SchartmannAA05}, SKIRTOR \citep{StalevskiMNRAS12,StalevskiMNRAS16,YangMNRAS20}, and ADAF \citep{ADAF}. Since \texttt{[skirtor2016]} includes more refined geometry settings and polar dust, we build the incident radiation field based on the latter. Of the three options in \texttt{[skirtor2016]}, ADAF is mainly developed for low-luminosity AGNs, and simulations based on SKIRTOR and Schartmann yield almost identical results. Considering that Schartmann performs better for high-redshift galaxies \citep{MountrichasAA21} and to reduce the size of the database for faster download and operation, we therefore implemented the Schartmann model only.

The shape of the photoionizing radiation field given by the Schartmann model in \texttt{[skirtor2016]} is described by a broken power-law \citep{SchartmannAA05, Yang2022}:
\begin{equation}
   \begin{array}{lll}
     \lambda L\mathrm{_\lambda} &\propto \lambda^{2} & ~ 8~\mathrm{nm} \leq\lambda \leq 50~\mathrm{nm} \\
     &\propto \lambda^{0.8} & ~50~\mathrm{nm} \leq\lambda \leq 125~\mathrm{nm} \\
     &\propto \lambda^{-0.5+\delta} & ~125~\mathrm{nm} \leq\lambda \leq 10~\mathrm{\mu m} \\
     &\propto \lambda^{-3} & ~10~\mathrm{\mu m} \leq\lambda \leq 1000~\mathrm{\mu m}. \\       
   \end{array}
\label{Schartmann_field}
\end{equation}

The blue solid line in Fig.~\ref{Fig:incident radiation field} displays an example of the disk continuum given by \texttt{[skirtor2016]}. The $\mathrm{\delta}$ parameter offers a degree of flexibility in the shape of the field between 125~$\mathrm{nm}$ and 10~$\mathrm{\mu m}$. In our model, $\mathrm{\delta}$ has three optional values: $-0.5$, $0.0$, and $0.5$. Considering the ionization potentials of most important atomic and ionic species corresponds to wavelengths below 100~$\mathrm{nm}$, $\mathrm{\delta}$ has minimal impact on most emission lines. It can only have a very weak indirect effect on the emission lines by affecting the temperature of gas and dust. However, it may have a more significant impact on the low-ionization emission lines in PDRs.

\subsubsection{The corona emission}
\label{sec: model/inciden radiation field/corona}

To model the X-ray coronal emission, the \texttt{[X-ray]} module of CIGALE provides two options: \texttt{[yang20]} \citep{YangMNRAS20} and \texttt{[lopez24]} \citep{ADAF}. The latter is developed primarily for low-luminosity AGNs, so we only adopted \texttt{[yang20]}.

\citet{YangMNRAS20} used a power-law with an exponential cutoff to describe the X-ray continuum from the corona region:
\begin{equation}
f_{\nu} \propto \mathrm{E}^{-\Gamma+1}\exp(-\mathrm{E}/\mathrm{E_{cut}}),
\label{eq:X_field}
\end{equation}
where $\mathrm{\Gamma}$ is the photon index, $\mathrm{E_{cut}}$ is the high-energy cut-off of the power-law, and $\mathrm{f_{\nu}}$ is scaled to the appropriate level through the UV-to-X-ray spectral slope $\mathrm{\alpha_{OX}}$, i.e.
\begin{equation}
\mathrm{\alpha_{OX}} = -0.3838 \log(\mathrm{L_{2500 \text{\AA}}}/\mathrm{L_{2keV}}),
\label{eq:aox}
\end{equation}
In the \texttt{[yang20]} module, $\mathrm{\Gamma}$ and $\mathrm{E_{cut}}$ are both free parameters, with default values of 1.8 and 300 keV, respectively, and the value of $\mathrm{\alpha_{OX}}$ ranges from $-1.9$ to $-1.1$, with a step size of 0.1. \citet{YangMNRAS20} noted that for the typical value from observations of Seyfert galaxies \citep{Dadina2008, Ricci_2017}, detailed fitting of the X-ray spectrum found $\mathrm{\Gamma} \approx 1.8$ \citep{Yang_2016, LiuTeng, Ricci_2017}. Another commonly used broken power-law model \citep{XSPEC} of corona divides the spectrum into a hot corona ($\gtrsim$ 2 keV) and a warm corona ($\textless$ 2 keV). The typical photon index of the hot corona is about 1.9, while the typical value of the warm corona is around 2.4 due to the excess of soft X-rays \citep{Porquet2004}. In recent years, an increasing number of objects with $\mathrm{\Gamma} > 2.5$ have been reported \citep{Iwasawa2024, Wolf2023}, with some of them reaching very high values of 4.7 to 5 \citep{jiang2025, Sacchi2022}. For AGN emission lines, the most important role is played by the soft X-rays from the warm corona, as they provide photons with energies comparable to the ionization potentials of most emitting species, thereby driving the ionization and excitation of the gas. However, \texttt{[yang20]}, as a module specifically developed for fitting X-ray data, is mainly designed to describe the hard X-ray emission from the hot corona. To ensure the consistency of the models used in CIGALE, we developed \texttt{[nebular\_AGN]} based on \texttt{[yang20]}. However, this design choice also introduces a limitation when hard X-ray data (above 2 keV) and emission-line fluxes are fitted simultaneously using both \texttt{[yang20]} and \texttt{[nebular\_AGN]}. The emission lines are primarily sensitive to the soft X-ray continuum from the warm corona and therefore favour a steeper photon index (soft X-ray excess), whereas the hard X-ray spectrum dominated by the hot corona typically shows a flatter photon index ($\Gamma \lesssim 2$). Since \texttt{[yang20]} adopts a single photon index for the entire X-ray continuum, the emission-line and hard X-ray data may favour inconsistent $\Gamma$ values, potentially biasing the fitting results. More accurate modelling of AGN emission lines will require future models that better describe the warm corona.

Since $\mathrm{E_{cut}}$ is higher than the highest observable energy of most X-ray observatories, its effect on the simulations is negligible. Therefore, we fixed it at 300 keV in the incident radiation field for our simulation, regardless of the value adopted in \texttt{[yang20]}.

To limit the size of the database of the \texttt{[nebular\_AGN]} module for efficient downloading and fitting, while matching the observations as closely as possible, we set six options for $\mathrm{\Gamma}$: 1.8, 2.4, 3.0, 3.6, 4.2, and 4.8, and three options for $\mathrm{\alpha_{OX}}$: $-1.9$, $-1.5$, and $-1.1$, to build a parameter grid for the incident radiation field in our simulation. When $\mathrm{\Gamma}$ and $\mathrm{\alpha_{OX}}$ are set to other values in \texttt{[yang20]}, we adopt a nearest-neighbour approach in the \texttt{[nebular\_AGN]} module, mapping them to the closest values on the grid rather than interpolating between grid points, in order to avoid introducing poorly constrained intermediate spectra, given the strong non-linearity of emission-line responses to the incident SED shape. In Fig.~\ref{Fig:incident radiation field}, we use the purple solid line to show an example of the corona continuum given by \texttt{[yang20]}. 

\subsubsection{Fixing the discontinuity}
\label{sec: model/Fixing the discontinuity}

Because the \texttt{[X-ray]} module and the \texttt{[AGN]} module were developed separately, each focusing on photometric fitting within its own wavelength range, the corona and accretion disk emissions were not treated as a single, unified incident radiation field responsible for exciting AGN broad and narrow emission lines in CIGALE. Consistency between the two modules is specified only by the UV-to-X-ray spectral slope $\mathrm{\alpha_{OX}}$. As a result, the continuity of the shape of the SED between the X-ray and UV regimes is not guaranteed. As illustrated in Fig.~\ref{Fig:incident radiation field}, the CIGALE coronal emission cuts off at 5~nm, while the accretion disk emission cuts off at 8~nm. In addition, the luminosity of the Schartmann model decreases rapidly at wavelengths shorter than 50 nm. Together, these effects introduce a significant and unphysical discontinuity in the 5–50~nm spectral range, which encompasses the ionization potentials of the atomic and ionic species responsible for the key emission lines, as summarised in Table~\ref{Table:IE} and indicated by the coloured vertical dash-dotted lines in Fig.~\ref{Fig:incident radiation field}.

\begin{table}[h!]
\caption{Ionization energies and corresponding wavelengths of the main emitting species.}
\tiny
\centering                         
\begin{tabular}{lcc}       
\hline
element &  IE (eV) & $\lambda_0$ (nm) \\
\hline
He$^+$ & 54.42 & 22.78 \\
C$^{2+}$ & 47.89 &25.89 \\
O$^{+}$ & 35.12 & 35.30 \\
He & 24.59 & 50.42 \\
C$^+$ & 24.38 &50.85 \\
N &  14.53 & 85.32\\
O & 13.62 & 91.03 \\
H  & 13.60 & 91.16 \\
S & 10.36 & 119.68\\
Mg & 7.65 & 162.07 \\
\hline                                                           
\end{tabular}         
\label{Table:IE}     
\end{table}

According to Kramers’ approximate formula, the photoionization cross-section is proportional to the cube of the incident photon wavelength for wavelengths shorter than the corresponding ionization threshold wavelength. Therefore, the discontinuity in the 5-50 nm range will affect the ion abundances and their distribution.

\begin{figure}[h!]
\centering
\includegraphics[width=8cm]{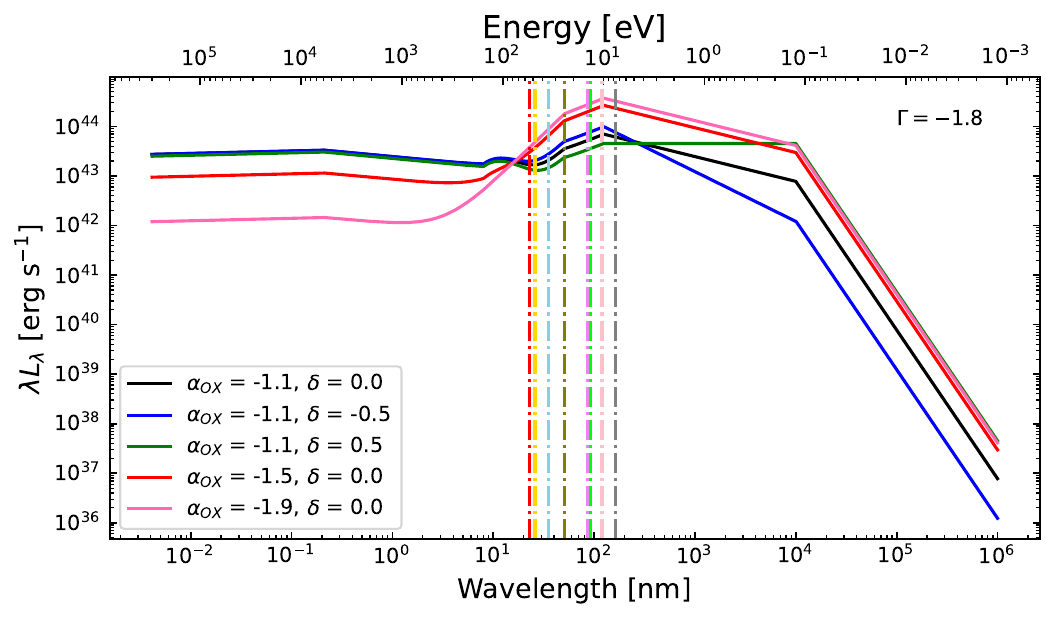}
\caption{Spectral shape of the incident radiation field for different $\alpha_{OX}$ and $\delta$ normalised to a bolometric luminosity of $10^{45}$ erg s$^{-1}$. The coloured vertical dotted lines indicate the ionization potentials of different species and are the same as in Fig.~\ref{Fig:incident radiation field}.}
\label{Fig:aox and delta}
\end{figure}

Fixing this discontinuity is challenging because X-ray observatories, which can observe soft X-ray such as Chandra and XMM–Newton, do not operate at sufficiently low energies to directly constrain this spectral region, due to both instrumental limitations and absorption by the Galactic foreground. As a result, the shape of the spectrum in this regime remains poorly constrained observationally. For a long time, most discussions on soft X-ray models were limited to above 0.2~keV \citep{Laor_1997}, and the connection between the X-ray and UV spectra has typically been characterised only through the X-ray–to–UV spectral slope, $\mathrm{\alpha_{OX}}$. \citet{Timlin21} used He$^+$ to trace the spectral range from ultraviolet (UV, 150-250~nm), through the extreme ultraviolet (EUV, 5-30~nm), to the X-ray regime (2~keV), and found evidence of an unknown physical mechanism—independent of redshift—that couples the emission from the X-ray-emitting hot corona, the UV-emitting accretion disk, and the EUV-emitting inner regions of the accretion disk. This mechanism appears to regulate the overall continuum shape from the UV to the X-ray regime. \citet{Jiang_2025} used 3D radiation magnetohydrodynamic simulations to investigate the physical origin of the spectrum from EUV to soft X-rays and found that when the accretion rate is close to the Eddington value, a black hole with a mass of $\mathrm{10^{8}}$ $\mathrm{M_\odot}$ can form a power-law spectrum varying between $\mathrm{\nu^{-1}}$ and $\mathrm{\nu^{-3}}$ in the energy range of about 0.01–1~keV, consistent with observations at energies $\gtrsim$ 0.2~keV.

Considering that a first-principles model capable of consistently connecting the EUV and soft X-ray regimes is still missing, we extended the power-law model used by \citet{YangMNRAS20} to 8~nm and then added a normalised blackbody spectrum beyond 8~nm, combined with the Schartmann disk spectrum extended down to 0~nm, to ensure continuity of the incident radiation field. For most combinations of the spectral parameters $\mathrm{\Gamma}$ and $\mathrm{\alpha_{OX}}$, we set the temperature of the blackbody radiation to $3.3 \times 10^{5}$~K to obtain a reasonably smooth shape. However, for a small number of those spectral parameter combinations, this choice led to unphysical convexities or concavities in the spectrum. In these cases, we adopted a temperature of $1 \times 10^{6}$~K. This component may correspond to EUV emission from the inner regions of the accretion disk as described by \citet{Timlin21}, or a so-called “cold corona”. However, we emphasise that this is a practical solution within the CIGALE framework, not a physical truth. In Fig.~\ref{Fig:incident radiation field}, the coronal continuum corrected with an added blackbody component is shown as the cyan dashed line, while the corresponding disk continuum extended to 0~nm is represented by the green dashed line. Their sum, shown as the black solid line, defines the final spectral shape adopted as the incident radiation field.

In Fig.~\ref{Fig:aox and delta}, we illustrate the impact of varying $\alpha_{OX}$ and $\delta$ on the shape of the incident radiation field. The effect of varying $\Gamma$ on the incident radiation field is shown in Fig.~\ref{Fig:Feltre incident radiation field}. It should be noted that although we have attempted to fix the discontinuity between the X-ray and UV parts of the incident radiation field, the extreme parameter combinations with $\mathrm{\alpha_{OX}}$ = -1.1 and $\Gamma \geq$ 4.2 still introduce a significant and physically unrealistic bump around 10 nm. As discussed in Sect.~\ref{sec: benchmark/BPT}, compared to the cases with $\Gamma$=3.0, such extreme values provide only a marginal increase in the coverage of our simulated diagnostic grids in the BPT/VO87 diagrams. We therefore advise caution when using these parameter combinations and do not recommend them for general fitting applications. In practice, such extremely high values of $\Gamma$ are very rare.

\subsubsection{Other sources of ionization}
\label{sec: model/inciden radiation field/other}

The cosmic ray background is included as an additional source of ionizing radiation at redshift $z=0$, although it has a negligible influence compared to photons in fully ionized regions. We decided not to include outflows in our model. If the mechanical energy input in the ISM by outflows can be non-negligible in some specific viewing-angle configurations, their effects are highly geometry-dependent. We considered that including them as an additional source of ionization could blur the effects of more dominant parameters, exacerbate the degeneracy among parameters and increase the complexity and computational cost of the simulations.

\subsection{The elemental abundances and dust}
\label{sec: model/abundance}
We followed the elemental abundance prescriptions described by \citet{TheuleAA24}. Briefly, in our simulations, the model adopts the cosmic abundance standard and scaling developed by \cite{NievaPrzybillaAA12}, based on the observed metallicities of 29 early B-type stars in the local Galactic region rather than on solar abundance standards \citep{AsplundARAA09,GrevesseASS10,LoddersASSP10}. At the so-called local Galactic concordance, 12+log(O/H)$_{GC}$=8.76, which is close to the primordial solar abundance of 8.73 estimated by \citet{AsplundARAA09} and \citet{LoddersASSP10}, and (O/H)$_{GC}$=5.76$\times$10$^{-4}$, (N/H)$_{GC}$=6.17$\times$10$^{-5}$, and Z$_{GC}$=0.01425. For user convenience, and to maintain consistency with the gas metallicity scale used in the \texttt{[nebular]} module, we adopted the gas metallicity as the metallicity scale in the \texttt{[nebular\_AGN]} module. All available abundance options can be found in Table~1 of \citet{TheuleAA24}.

Dust grains in the model have a dual impact on the NLR and BLR emission spectra: They deplete the ISM from refractory elements (cooling agents) and cause wavelength-dependent absorption and scattering of the incoming light. We chose grains with an appropriate size distribution and abundance given by the default settings in \texttt{Cloudy v23.01} to reproduce the overall observed extinction properties for the ISM of the Milky Way in our gas cloud model. This grain distribution includes both a graphitic and a silicate component, and the ratio of extinction to reddening is $\mathrm{R_V \equiv A_V/E(B-V) = 3.1}$ \citep{CLOUDY}. The abundance of dust is scaled as the metallicity changes. We only added the dust component to the NLRs and not to the BLRs because the high temperatures in the BLRs cause the dust grains to sublime. Consequently, metallicities are taken without the depletion factor in BLRs, while depletion is considered in NLRs. 

We note that the dust discussed here is different from the polar dust introduced in Section~\ref{sec: model/geometry}. The dust in this section is part of the NLR structure, whereas the polar dust, according to our geometrical model, is assumed to be distributed outside the NLRs and attenuates the emission originating from the NLRs and BLRs. The properties of polar dust are specified by the \texttt{[skirtor2016]} module.

\subsection{Integration of the model data into CIGALE}
\label{sec: model/integration}
After defining the model geometry, the incident radiation field, and the dust and elemental abundance settings, we generated a total of 281,232 models corresponding to different combinations of parameter values. For each model, the \texttt{Cloudy v23.01} photoionization simulation generated the corresponding continuum and emission-line spectra, serving as the spectral templates in the database of the \texttt{[nebular\_AGN]} module. Since the shape of the incident radiation field is specified by parameters defined in \texttt{[skirtor2016]} and \texttt{[yang20]}, both modules are required for running \texttt{[nebular\_AGN]}.

In Table \ref{Table:parameters}, we summarise the names of all the parameters and their available values in the \texttt{[nebular\_AGN]} module. We also list six additional parameters that are important for the \texttt{[nebular\_AGN]} module but are specified in other modules of CIGALE. As mentioned in Section~\ref{sec: model/inciden radiation field}, $\mathrm{\Gamma}$ and $\mathrm{\alpha_{OX}}$ are defined in [\texttt{yang20}], while $\mathrm{\delta}$ is defined in [\texttt{skirtor2016}]. Among these, $\mathrm{\Gamma}$ and $\mathrm{\delta}$ are free parameters, whereas $\mathrm{\alpha_{OX}}$ has nine optional values, but only three of them are supported in \texttt{[nebular\_AGN]}. Here, we only list the options of these parameters supported by [\texttt{nebular\_AGN}]; any other values provided as input are automatically mapped to the closest optional value in the [\texttt{nebular\_AGN}]. The parameter \texttt{fracAGN} in [\texttt{skirtor2016}] is a free parameter that sets the fractional contribution of the AGN to the total infrared luminosity of the galaxy by default and thus directly affects the emission strength of both the NLR and the BLR. Since users can change the wavelength range used to calculate \texttt{fracAGN}, it is not always limited to infrared. The parameter \texttt{i} represents the inclination angle, while \texttt{oa} denotes the angle measured from the equatorial plane to the outer edge of the torus. $90^\circ - \mathrm{oa}$ corresponds to the half-opening angle of the cone attenuated only by polar dust, i.e. the region unobscured by the dust torus. The combination of \texttt{oa} and the viewing angle \texttt{i} determines the type of AGN: For \texttt{i} $\in [0,\, 90^\circ - \mathrm{oa})$, the system is observed face-on (type I); for \texttt{i} $\in [90^\circ - \mathrm{oa},\, 90^\circ]$, it corresponds to an edge-on (type II) view. When \texttt{i} and \texttt{oa} correspond to a type II AGN, the contribution from the BLR is set to zero in the [\texttt{nebular\_AGN}] module.

\section{Model benchmark}
\label{sec: benchmark}

In this section, we benchmark our \texttt{[nebular\_AGN]} module by (\textit{i}) fitting the photometric and emission-line data of the composite quasar spectrum of \citet{VandenBerkAJ01} using CIGALE with the \texttt{[nebular\_AGN]} module, (\textit{ii}) comparing the metallicity predictions of \texttt{[nebular\_AGN]} with the empirical calibration based on emission-line ratios, (\textit{iii}) comparing the simulated emission-line ratios generated by \texttt{[nebular\_AGN]} with observations in the BPT/VO87 diagrams \citep{BPT, Veilleux1987, Kewley2001, KauffmannMNRAS03a} using X-ray-selected AGN samples and SDSS objects, and (\textit{iv}) directly comparing the quality of key emission-line fitting with and without the \texttt{[nebular\_AGN]} module. All the CIGALE simulations mentioned in this section use the following combination of modules: \texttt{[sfhdelayed]}, \texttt{[bc03]}, \texttt{[nebular]}, \texttt{[dustatt\_modified\_starburst]}, \texttt{[dl2014]}, \texttt{[skirtor2016]},  \texttt{[yang20]}, \texttt{[nebular\_AGN]}, and \texttt{[redshifting]}.

\subsection{Benchmark on a quasar composite spectrum}
\label{sec: benchmark/quasar}
\citet{VandenBerkAJ01} presented a widely used composite quasar spectrum constructed by stacking 2200 observed quasar spectra from the Sloan Digital Sky Survey (SDSS) covering an observed wavelength range of 380--920~nm at a spectral resolution of 1800. The quasar sample spans a redshift range of 0.044--4.789 with an absolute r-band magnitude range from $-26.5$ to $-18.0$. This composite spectrum contains many prominent AGN emission lines, such as \ion{C}{iv}, \ion{C}{iii}], \ion{Mg}{ii}, [\ion{O}{ii}], [\ion{O}{iii}], and [\ion{N}{ii}]. These lines provide an excellent benchmark for evaluating the fitting performance of our new module. We used the composite spectrum rather than individual quasar spectra for testing, as it covers a broad redshift range and includes nearly all major UV-optical emission lines. Built from the stacking of over two thousand quasars, it provides a more representative and general spectral template than the spectrum of any individual object. Individual quasar spectra may suffer from selection effects and intrinsic peculiarities, and are further limited by instrumental wavelength coverage, which typically allows only a subset of emission lines to be observed.

Before using the quasar composite spectrum as a benchmark, several preprocessing steps are required. The composite spectrum is provided in a normalised form; therefore, we assumed a bolometric luminosity of $10^{46}\,\mathrm{erg\,s^{-1}}$ and scaled the spectrum accordingly. We set the redshift of the spectrum to $z=1.25$, corresponding to the average redshift of the 2200 quasars reported by \citet{VandenBerkAJ01}. We assumed a signal-to-noise ratio (S/N) of 5 and performed a Monte Carlo simulation, adopting the corresponding uncertainty of 20\% to perturb the spectrum and generate 100 artificial realisations. The standard deviation at each data point was then taken as the corresponding noise.

From this processed spectrum, we derived the corresponding photometric fluxes in the SDSS $u$, $g$, $r$, $i$, $z$ bands and the 2MASS $J$ band together with their associated uncertainties. We also measured the fluxes of 24 sets of emission lines using a multi-Gaussian fitting approach; these lines are indicated by black dashed lines in Fig.~\ref{Fig:line profile}, and their corresponding uncertainties are provided as well. We then used these measurements as input data for CIGALE, employing the \texttt{[nebular\_AGN]} module for spectral fitting. In setting the parameter space, we fixed \texttt{fracAGN} = 0.99 and \texttt{i} = 0 in the \texttt{[skirtor2016]} module to ensure a quasar-like configuration. For the \texttt{[yang20]} module, we explored \texttt{gam} = 1.8, 2.4, 3.0 and \texttt{alpha\_ox} = -1.1, -1.5, -1.9. In the \texttt{[nebular\_AGN]} module, we varied the \texttt{metallicity} over 0.001, 0.005, 0.011, 0.014, 0.019, 0.033, 0.05; the NLR and BLR covering factors over \texttt{f\_NLR} = 0.1, 0.2, 0.3 and \texttt{f\_BLR} = 0.1, 0.2, 0.3; and the ionization parameters over \texttt{logU\_NLR} = -3.8, -3.5, -3.0, -2.5, -2.0, -1.8, -1.5, -1.2 and  \texttt{logU\_BLR} = -3.0, -2.5, -2.0, -1.8, -1.5, -1.2. The line widths were fixed to 600~km\,s$^{-1}$ for the NLR and 3000~km\,s$^{-1}$ for the BLR. We also adjusted the e-folding time and the age of the stellar population in the \texttt{[sfhdelayed]} module to ensure that the stellar population age did not exceed the cosmic age at $z=1.25$, although these parameters do not contribute to the emission-line modelling in this work. All other parameters were set to their default values.

We obtained the following parameters for the best-fit model: \texttt{gam} = 3.0, \texttt{alpha\_ox} = -1.5, \texttt{metallicity} = 0.05, \texttt{logU\_NLR} = -3.0, \texttt{logU\_BLR} = -3.0, \texttt{f\_NLR} = 0.1, and \texttt{f\_BLR} = 0.2. The measured band fluxes and emission-line fluxes, together with the corresponding best-fit results, are listed in Table~\ref{Table:VenDenBerkFit}. Since the uncertainties are artificially generated through Monte Carlo simulations, the normalised residuals of the fit are meaningless for the fitting-quality evaluation. Instead, we assess the fitting quality using the relative error (RE) between the observed and fitted values. Fits with RE $\leq 0.5$ are considered reasonably acceptable. A direct comparison between the best-fit spectrum and the input quasar composite spectrum is shown in Fig.~\ref{Fig:line profile}.

For the photometric data, all bands except the SDSS $u$ band exhibit RE values below 0.1, indicating an excellent overall fitting quality. The relatively poor performance in the $u$ band (RE = 0.17) arises from the steep decline of the Schartmann continuum in the UV regime, as illustrated in Fig.~\ref{Fig:line profile}.

\begin{table}[htbp]
\caption{Evaluation of fitting results.}
\centering
\begin{tabular}{|l|l|l|l|}
\hline
 data & band flux & best-fit & RE \\
\hline
 SDSS.u & 70.41 $\pm$ 0.56 & 58.59 & 0.17 \\
 SDSS.g. & 73.55 $\pm$ 0.40 & 74.00 & 0.01\\
 SDSS.r & 98.14 $\pm$ 0.64 & 94.69 & 0.04\\
 SDSS.i & 103.33 $\pm$ 0.67 & 106.95 & 0.03\\
 SDSS.z & 105.66 $\pm$ 0.53 & 115.50 &  0.09\\
 2MASS.J & 133.23 $\pm$ 0.76 & 141.86 & 0.06\\
 \hline
  data & line flux & best-fit & RE \\
 \hline
 Lyman $\alpha$ & 271.40 $\pm$ 22.57 & 156.00 & 0.43\\
\,\ion{C}{iv}\,1548,1551 & 123.33 $\pm$ 7.38 & 9.08 & 0.93 \\
\,\ion{He}{ii}\,1640 & 9.59 $\pm$ 2.55 & 7.80 & 0.19\\
\,\ion{O}{iii}]\,1661, 1666 & 3.93 $\pm$ 2.60 & 4.54 & 0.15\\
\,\ion{Al}{iii}\,1855,1863 & 8.35 $\pm$ 2.64 & 1.07 & 0.87\\
\,\ion{C}{iii}]\,1907,1909 & 28.38 $\pm$ 3.07 & 18.42 & 0.35\\
\,\ion{Mg}{ii}\,2796,2803 & 71.95 $\pm$ 3.37 & 43.81 & 0.39\\
\,[\ion{O}{ii}]\,3727,3729 & 3.23 $\pm$ 0.64 & 1.60 & 0.50\\
\,[\ion{Ne}{iii}]\,3869 & 2.53 $\pm$ 0.67 & 0.56 & 0.78\\
\,\ion{He}{i}\,3889 & 0.63 $\pm$ 0.51 & 0.88 & 0.40\\
\,[\ion{Ne}{iii}]\,3967 & 0.52 $\pm$ 0.49 & 0.17 & 0.68\\
\,H$\delta$ & 4.85 $\pm$ 1.49 & 2.24 & 0.54\\
\,H$\gamma$ & 8.43 $\pm$ 1.09 & 4.43 & 0.47\\
\,[\ion{O}{iii}]\, 4363 & 1.08 $\pm$ 0.57 & 0.06 & 0.94\\
\,H$\beta$ & 20.69 $\pm$ 2.00 & 13.16 & 0.36\\
\,[\ion{O}{iii}]\,4959 & 2.54 $\pm$ 0.67 & 3.42 & 0.35\\
\,[\ion{O}{iii}]\,5007 & 7.45 $\pm$ 1.06 & 10.25 & 0.38\\
\,\ion{He}{i}\,5876 & 1.81 $\pm$ 0.71 & 2.02 & 0.12\\
\,[\ion{O}{i}]\,6300 & 0.58 $\pm$ 0.73 & 2.59 & 3.50\\
\,H$\alpha$ & 98.59 $\pm$ 3.93 & 142.47 & 0.45\\
\,[\ion{N}{ii}]\,6548 & 1.66 $\pm$ 1.47 & 1.5 & 0.11\\
\,[\ion{N}{ii}]\,6583 & 6.03$\pm$ 1.38 & 4.35 & 0.28\\
\,[\ion{S}{ii}]\,6716 & 1.69 $\pm$ 0.46 & 0.93 & 0.45\\
\,[\ion{S}{ii}]\,6731 & 1.50 $\pm$ 0.43 & 1.24 & 0.17\\
\hline
\end{tabular}
\tablefoot{The band fluxes and emission-line fluxes used for the fitting, the best-fit results, and the RE for assessing the fitting quality are listed. The band fluxes are expressed in units of $\mu$Jy, while the emission-line fluxes are given in units of $10^{-19}\ \mathrm{W\,m^{-2}}$.}
\label{Table:VenDenBerkFit}  
\end{table}

Among the 24 sets of emission-line flux measurements, 17 sets have RE values $\leq 0.5$, corresponding to 71\% of the total. For the poorly fitted emission lines, the underestimation of the \ion{C}{iv}\,1548,1551 doublet can be partly attributed to the insufficient production of C$^{2+}$ ionizing photons under the combined parameter set of \texttt{gam}, \texttt{alpha\_ox}, and \texttt{logU\_BLR}. However, even the most extreme combinations of parameters do not produce sufficiently strong \ion{C}{iv} emission, suggesting that the primary limitation is the absence of shock ionization from AGN outflows. The under-prediction of [\ion{O}{iii}]\,4363 is likely due to the superposition effect of the composite spectrum. Seyfert galaxies typically exhibit [\ion{O}{iii}]\,4363/[\ion{O}{iii}]\,5007 < 0.1, while the rare cases with ratios exceeding 0.1 are generally associated with log\,($\mathrm{nH \, [cm^{-3}]})$ > 6 \citep{Baskin2005b}. Such a high density lies outside the parameter space covered by our models and exceeds the critical densities of many forbidden lines. In this composite spectrum, [\ion{O}{iii}]\,4363/[\ion{O}{iii}]\,5007 is approximately 0.145. Therefore, it can be inferred that, under realistic AGN conditions, such a high [\ion{O}{iii}] line ratio is unlikely to coexist with prominent [\ion{S}{ii}], [\ion{O}{ii}], and [\ion{N}{ii}] emission lines observed in composite quasar spectra. Consequently, we are unable to obtain a satisfactory estimate of [\ion{O}{iii}]\,4363 while simultaneously maintaining satisfactory fits to the other emission lines. The under-prediction of \ion{Al}{iii}~1855,1863 may be due to the same reason. The over-prediction of [\ion{O}{i}]\,6300 indicates that the model contains a more extended partially ionized zone (PIZ) than the actual system. Theoretically, since our model does not include the PDRs, a fraction of the [\ion{O}{i}]\,6300 emission is expected to be absent. However, in high-luminosity quasars, the gas clouds are likely matter-bounded. In contrast, we impose an ionization-front stopping criterion in our simulation, which ensures that the model always develops a substantial PIZ around the ionization front to produce [\ion{O}{i}]\,6300 emission. As a result, this line is overestimated in the best-fit result. The RE values for \ion{Ne}{iii}\,3869, 3967 and H$\delta$ are smaller than those of other poorly fitted emission lines. We attribute these discrepancies primarily to uncertainties in the continuum estimation over the rest-frame 360--450 nm wavelength range. As shown in Fig.~\ref{Fig:line profile}, this spectral region is heavily affected by severe line blending, and the best-fit continuum appears to be systematically overestimated.

We further check the best-fit line profiles in Fig.~\ref{Fig:line profile}. It is worth noting that the current version of CIGALE performs the fitting based on band fluxes and emission-line fluxes rather than through direct full-spectrum profile fitting. In addition, the line profiles in the composite spectrum are inevitably affected by spectral stacking effects. Therefore, a high-fidelity match between the line profiles of the best-fit and composite spectra is not expected. This figure is intended to illustrate the fitting details rather than to assess the overall fitting quality. We find that the profiles of most best-fit lines with RE $\leq 0.5$ are broadly consistent with the corresponding emission-line profiles in the composite quasar spectrum. Even in the wavelength range of 360--450 nm, where the continuum is overestimated in the best-fit spectrum and direct comparison is difficult, the line profiles are broadly consistent after continuum subtraction. The only notable exception is Ly$\alpha$. Even though a broad-line width of 3000 km s$^{-1}$ is adopted, the best-fit Ly$\alpha$ profile remains significantly narrower than that in the composite quasar spectrum. Given that Ly$\alpha$ emission often exhibits complex profiles due to outflows, strong resonant scattering, and IGM absorption, we attribute this discrepancy primarily to stacking-induced broadening and profile distortion in the composite spectrum rather than to a limitation of the model itself.

\subsection{Benchmark on empirical metallicity diagnostics}
\label{sec: benchmark/Dors}

In this section, taking metallicity as an example, we compare the metallicity given by the \texttt{[nebular\_AGN]} module with that obtained from an empirical line-ratio diagnostic, in order to assess the reliability of \texttt{[nebular\_AGN]} in deriving physical parameters. \citet{Flury2020} developed the first AGN calibration anchored to $T_e$-based O/H abundances rather than photoionization-model abundances, providing a single equation to estimate AGN metallicity directly from the [\ion{O}{iii}]/H$\beta$ and [\ion{N}{ii}]/H$\alpha$ ratios. \citet{Dors2020} showed that the discrepancy between $T_e$-based and photoionization-model-based estimates of O/H abundances in Seyfert 2 NLRs mainly arises from the inappropriate application of the $t_2$--$t_3$ relation, originally established for \ion{H}{ii} regions, to AGNs. Based on \texttt{Cloudy} photoionization models, \citet{Dors2020} derived a new $t_2$--$t_3$ relation for AGNs, which formed the basis of the direct-method metallicity calibration presented by \citet{Dors2021}.

We adopted the direct method derived by \citet{Dors2020,Dors2021} based on a sample of 56 Seyfert 1 and 35 Seyfert 2 galaxies to benchmark our module. The calibration is expressed as
\begin{equation}
12+\log(\mathrm{O}/\mathrm{H}) = (-1.00 \pm 0.09)P + (0.036 \pm 0.003)R_{23} + (8.80 \pm 0.06)
\label{eq:Dors_metallicity},
\end{equation}
where $R_{23}=([\ion{O}{ii}]\,3727,3729+[\ion{O}{iii}]\,4959,5007)/\mathrm{H\beta}$, and $P=([\ion{O}{iii}]\,4959,5007/\mathrm{H\beta})/R_{23}$ is an indicator of the hardness of the ionizing radiation and is used to account for the effect of ionization conditions on $R_{23}$. The validity range of this calibration is $8.0 < 12 + \log(\mathrm{O}/\mathrm{H}) < 9.2$, corresponding to $0.003 < \mathrm{Z_{gas}} < 0.045$ in CIGALE.

We used CIGALE with the \texttt{[nebular\_AGN]} module, adopting the \texttt{savefluxes} mode (i.e. a mode that does not fit observational data but instead generates simulated spectra solely based on the input parameters) to produce a series of simulated spectra corresponding to different metallicities. We then used the line fluxes of [\ion{O}{ii}]\,3727,3729, [\ion{O}{iii}]\,4959,5007, and H$\beta$ from these simulated spectra to compute $R_{23}$ and $P$. Subsequently, these values were substituted into Eq.~\ref{eq:Dors_metallicity} to derive the metallicity using the calibration of \citet{Dors2021}. Finally, we compared the \citet{Dors2021} metallicities with the input CIGALE metallicities to evaluate the performance of the \texttt{[nebular\_AGN]} module and to examine how different choices of other parameters affect the estimated results.

\citet{Dors2021} pointed out that the electron density $\mathrm{N_{e}}$ derived from the [\ion{S}{ii}]\,6716/[\ion{S}{ii}]\,6731 line ratio is approximately the same for Seyfert 1 and Seyfert 2 galaxies in the sample used to establish the empirical calibration. Therefore, it can be assumed that the emission lines considered in their analysis originate from gas with similar physical conditions in both Seyfert 1 and 2 galaxies, implying that the diagnostic is applicable to both types. However, given that [\ion{S}{ii}]\,6716,6731 are forbidden lines predominantly emitted from the NLR, we interpret the derived $\mathrm{N_{e}}$ as representative of the electron density in the NLRs of Seyfert 1 and Seyfert 2 galaxies. In fact, the electron densities of both the Seyfert 1 and Seyfert 2 samples derived by \citet{Dors2021} are consistent with typical values for the NLR ($10^{2} - 10^{3}\,\mathrm{cm^{-3}}$). Consequently, we treat this diagnostic as primarily tracing the NLR. Accordingly, when generating the simulated spectra, we consider only the type 2 AGN case.

\begin{figure*}[h!]
\centering
\includegraphics[width=16cm]{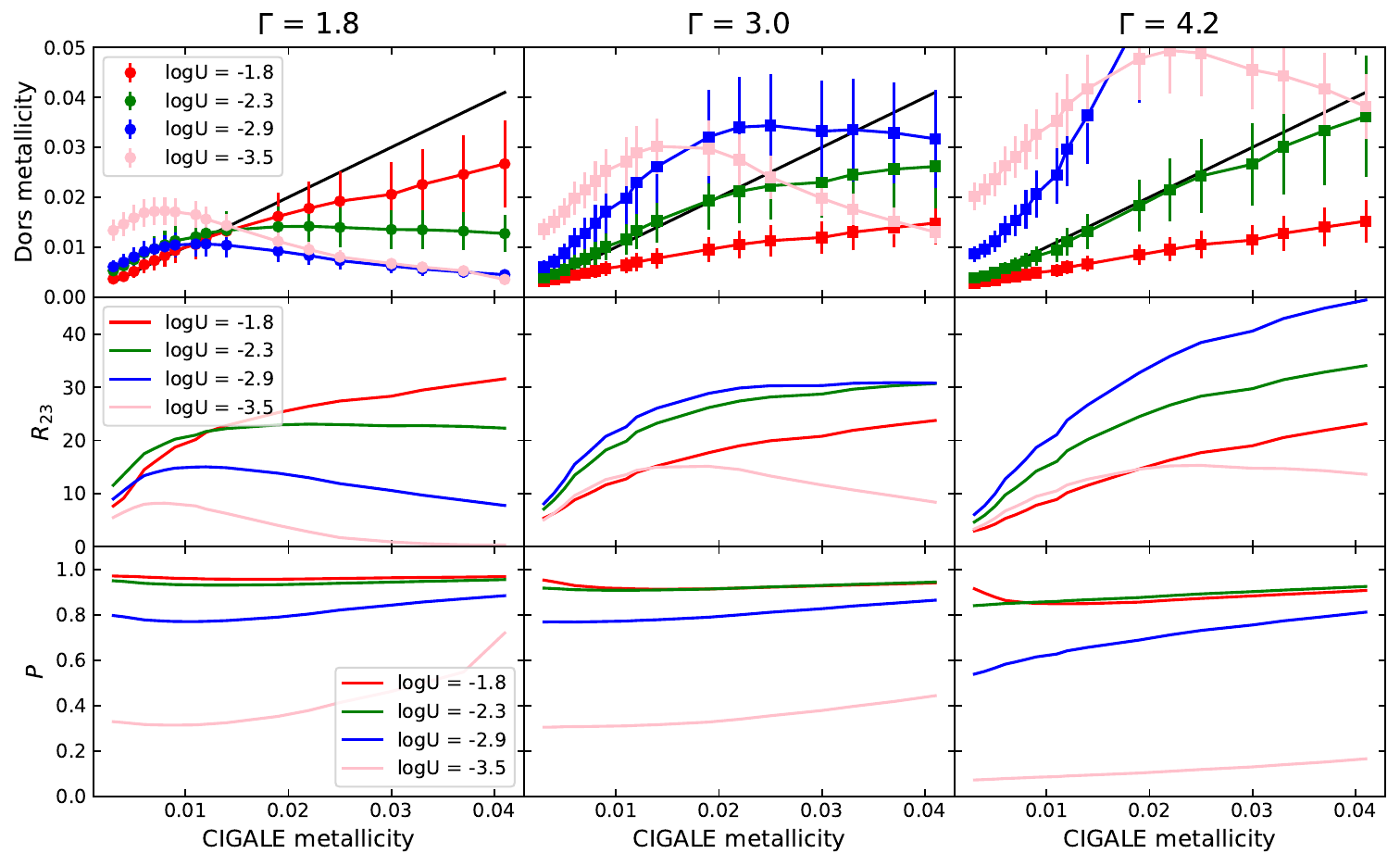}
\caption{Benchmark of metallicity predictions for models with different ionization parameters and photon indices. We fixed $\texttt{alpha\_ox} = -1.1$ and $\texttt{nH\_NLR} = 3.0$ and explored the agreement between the CIGALE metallicity and the \citet{Dors2021} metallicity under different combinations of $\texttt{gam}$ and $\texttt{logU\_NLR}$ as well as the corresponding trends of $R_{23}$ and $P$.}
\label{Fig:Dors_a}
\end{figure*}

\begin{figure*}[h!]
\centering
\includegraphics[width=16cm]{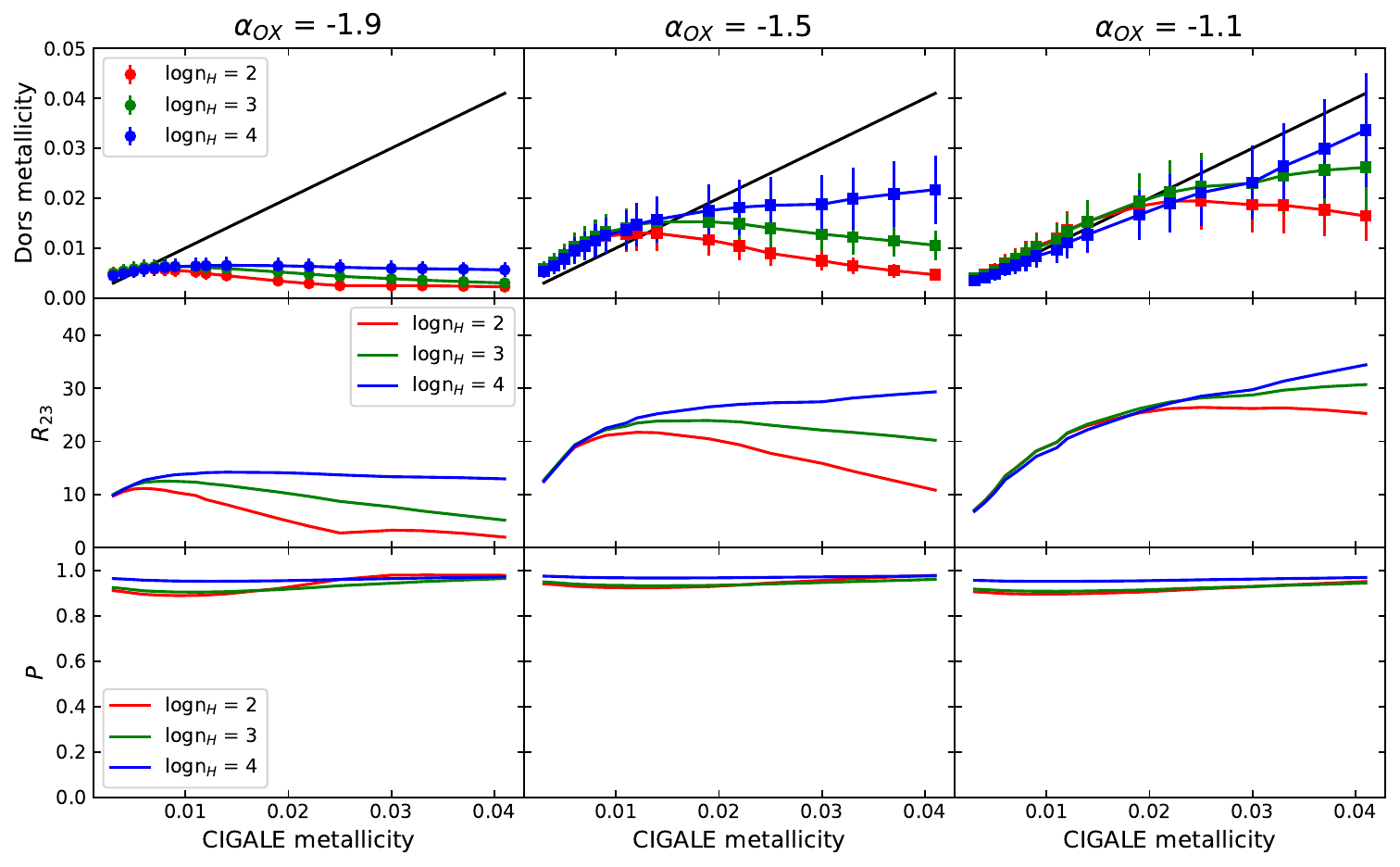}
\caption{Benchmark of metallicity predictions for models with different densities and UV-to-X-ray spectral slopes. We fixed $\texttt{gam} = 3.0$ and $\texttt{logU\_NLR} = 2.3$  and explored the agreement between the CIGALE metallicity and the \citet{Dors2021} metallicity under different combinations of $\texttt{alpha\_ox}$ and $\texttt{nH\_NLR}$ as well as the corresponding trends of $R_{23}$ and $P$.}
\label{Fig:Dors_b}
\end{figure*}

In defining the parameter space, we fixed \texttt{fracAGN} = 0.99 and \texttt{i} = 90 in the \texttt{[skirtor2016]} module to ensure a type 2 AGN configuration. For the \texttt{[yang20]} module, we explored \texttt{gam} = 1.8, 3.0, and 4.2, and \texttt{alpha\_ox} = -1.1, -1.5, and -1.9. In the \texttt{[nebular\_AGN]} module, we varied the \texttt{metallicity} over 0.003, 0.004, 0.005, 0.006, 0.007, 0.008, 0.009, 0.011, 0.012, 0.014, 0.019, 0.022, 0.025, 0.030, 0.033, 0.037, and 0.041; the hydrogen density \texttt{nH\_NLR} over 2.0, 3.0, and 4.0; and the ionization parameter of the NLR over \texttt{logU\_NLR} = -3.5, -2.9, -2.3, and -1.8. For the covering factors, \texttt{f\_NLR} was fixed to 0.1 and \texttt{f\_BLR} was set to 0.2. In practice, since the metallicity calibration of \citet{Dors2021} depends only on emission-line ratios, varying \texttt{f\_NLR} does not affect the results. In Figs.~\ref{Fig:Dors_a} and~\ref{Fig:Dors_b}, we compare the CIGALE metallicities of the simulated spectra with those derived from the \citet{Dors2021} calibration under different parameter settings, together with the corresponding values of $R_{23}$ and $P$.

Among the three parameters explored in Fig.~\ref{Fig:Dors_a}, log\textbf{U} and $\mathrm{\Gamma}$ mainly affect $R_{23}$ by regulating the ionizing photon flux, while metallicity influences $R_{23}$ in two aspects: the gas temperature and the abundances of species. From Fig.~\ref{Fig:Dors_a}, we find that for $\mathrm{\Gamma} = 1.8$, the case of log\textbf{U} = -3.5 yields metallicities derived from \citet{Dors2021} that are higher than the input CIGALE metallicities in the low-metallicity regime. In the other cases, the two metallicity estimates are broadly consistent when the CIGALE metallicity is below 0.014. However, once the CIGALE metallicity exceeds 0.014, all log\textbf{U} cases yield metallicities from \citet{Dors2021} that are systematically lower than the corresponding CIGALE values.

This trend arises from the coupled effects of the three parameters. In low-metallicity cases, the gas temperature is high due to the lack of efficient coolants, and the element abundances are low. Increasing log\textbf{U} raises the flux of oxygen-ionizing photons and thus significantly enhances $R_{23}$. However, for the log\textbf{U} = -1.8 case, $R_{23}$ instead decreases. We initially suspected that this was due to over-ionization, where O$^{+}$ and O$^{2+}$ are further ionized into higher states. However, inspection of the model shows that high-ionization species such as O$^{3+}$ increase only marginally, while the abundances of O$^{+}$ and O$^{2+}$ do not significantly decrease. The actual cause of the decrease in $R_{23}$ in the high-ionization cases lies in the stopping criterion of our model. To ensure that the \texttt{Cloudy} simulation stops on the ionization front, we adopt a stopping condition of $\frac{\mathrm{H^{+}}}{\mathrm{H^{0}}} = 0.01$. With a constant hydrogen density, this implies that the high-ionization models develop a larger ionized region before reaching this stopping condition, especially for low-metallicity cases, which do not have enough abundance of metals to consume the ionizing photons and cool the gas. Since dust is present in the NLR model, the emergent emission in high-ionization cases undergoes stronger dust attenuation due to the larger size of the model, leading to a decrease in $R_{23}$. Notably, although the log\textbf{U} = -3.5 case has the lowest $R_{23}$, its corresponding ionization hardness correction factor $P$ is very low. Consequently, the metallicity derived from the calibration of \citet{Dors2021} becomes higher than the CIGALE metallicity. A similar behaviour is observed in cases with $\mathrm{\Gamma}$ = 3.0 and 4.2, indicating a limitation of the correction capability of $P$ at low ionization parameters.

As metallicity increases, the gas temperature decreases. For $\mathrm{\Gamma} = 1.8$, the flux of oxygen-ionizing photons is relatively low, especially in the low-log\textbf{U} cases. Although the oxygen abundance increases with metallicity, the available ionizing photons are insufficient to maintain a high fraction of ionized oxygen species, which causes $R_{23}$ to decrease. Consequently, the metallicities estimated using the calibration of \citet{Dors2021} for log\textbf{U} = -2.3, -2.9, and -3.5 are all lower than the CIGALE metallicity. Only the log\textbf{U} = -1.8 case provides enough ionizing photons to ionize the increased amount of oxygen. In this case, the enhancement of emission lines due to higher metallicity outweighs the weakening effect caused by the reduction in collisional excitation efficiency due to lower temperatures, resulting in the best agreement between the \citet{Dors2021} and CIGALE metallicities among the $\mathrm{\Gamma} = 1.8$ cases.

For the $\mathrm{\Gamma}$ = 3.0 and 4.2 cases, the increase in $\mathrm{\Gamma}$ leads to a higher flux of oxygen-ionizing photons. At the low-metallicity end, the reduced abundance of metals allows a larger fraction of ionizing photons to ionize neutral hydrogen compared to the high-metallicity cases. This further enhances the increase in the ionized cloud size caused by the adopted stopping criterion, resulting in stronger dust attenuation and thus a reduction in $R_{23}$. At the high-metallicity end, these photons are consumed by the increased abundance of metals, the increased oxygen-ionizing photons allow more oxygen to be ionized, significantly enhancing $R_{23}$. The log\textbf{U} = -2.3 and -2.9 cases can ionize oxygen more efficiently, while being less affected by dust attenuation than the log\textbf{U} = -1.8 case, and therefore produce significantly higher $R_{23}$. In the log\textbf{U} = -2.3 case, when the CIGALE metallicity reaches 0.022, the weakening effect of emission lines due to the enhanced cooling from increasing metallicity and the enhancement of emission lines from more ionized species reach a balance, causing $R_{23}$ to remain nearly constant for metallicity $\geq 0.022$. Meanwhile, higher $\mathrm{\Gamma}$ also induces distortions in $P$ for the log\textbf{U} = -2.9 case. Therefore, for $\mathrm{\Gamma} = 3.0$ and $4.2$, only the log\textbf{U} = -2.3 case yields good agreement between the CIGALE metallicity and that from \citet{Dors2021}. This ionization parameter is close to the typical ionization parameter of AGN.

In Fig.~\ref{Fig:Dors_b}, since the density options in our module are generally below the critical density of the oxygen forbidden lines contributing to R23, $\log \mathrm{nH}$ mainly affects $R_{23}$ by regulating the gas temperature. Higher densities enhance collisional de-excitation in other forbidden-line transitions, weakening their emission and causing the gas to lose important cooling channels, thereby decreasing the cooling efficiency and increasing the equilibrium temperature. In the low-metallicity regime, the gas temperature is relatively high due to inefficient metal cooling; therefore, variations in $\mathrm{nH}$ have only a minor impact on $R_{23}$. However, at the high-metallicity end, the temperature decreases due to the presence of efficient coolants, and the temperature becomes more sensitive to changes in cooling efficiency induced by $\mathrm{nH}$, leading to an increase of $R_{23}$ with increasing $\mathrm{nH}$. On the other hand, $\mathrm{\alpha_{OX}}$ plays a role similar to $\Gamma$ in that both parameters regulate the flux of oxygen-ionizing photons, although $\Gamma$ mainly modifies the spectral slope while $\mathrm{\alpha_{OX}}$ changes the relative X-ray-to-UV normalisation.

Overall, we find that, for specific parameter combinations in the CIGALE \texttt{[nebular\_AGN]} module, our metallicity predictions are close to those of \citet{Dors2021}. For parameter combinations that yield metallicities inconsistent with those derived from the direct method, the discrepancies may arise from multiple factors. First, the recalibration of the $t_2$--$t_3$ relation by \citet{Dors2020, Dors2021} is also based on model grids generated with \texttt{Cloudy} photoionization simulations, but their adopted AGN incident radiation fields, parameter ranges, dust settings, and stopping criteria differ from those used in our models. Second, the calibration by $P$ cannot fully eliminate the effects of $\alpha_{OX}$, photon index, and ionization parameter, and the influence of hydrogen density is also not taken into account in the direct method. Given that the parameter space of real AGNs is evidently much narrower than that covered by the model grids, such discrepancies are expected. The Bayesian-like fitting approach adopted in CIGALE can mitigate the impact of the coverage of the model parameter space to some extent.

\subsection{The two galaxy test samples}
\label{sec: benchmark/samples}
We benchmark the simulated line ratios produced by the \texttt{[nebular\_AGN]} module against those observed in two samples. In this subsection, we present the selection process of the observational data.

\subsubsection{The OSSY sample}
\label{sec: benchmark/samples/OSSYsample}

The first sample was drawn from the Sloan Digital Sky Survey (SDSS) Data Release 7 \citep{Abazajian2009}. Using publicly available penalised pixel-fitting (pPXF) \citep{Cappellari2004} and Gas AND Absorption Line Fitting (GANDALF) IDL code \citep{Sarzi2006}, \citet{Oh2011} modelled the stellar and nebular components of SDSS spectra, based on the complete spectral map of galaxies with redshift z $\textless$ 0.2 in Data Release 7. They constructed a new database of absorption and emission line measurements, including a comprehensive list of recombination and collisional excitation lines, and provided both the fluxes and widths of these lines. The catalogue contains a total of 664,187 objects. By imposing an S/N threshold of 5 on the seven emission lines used in the BPT/VO87 diagrams, the sample was reduced to 285,631 objects including both AGN and non-AGN host galaxies.

\subsubsection{The X-ray-selected AGN sample}
\label{sec: benchmark/samples/X-ray selected samples}

The X-ray-selected sample was obtained through a positional cross-match between the 4XMM-DR9 catalogue of serendipitous X-ray sources \citep{WebbAA20} containing over 550,000 entries and the SDSS DR12 photometric catalogue \citep{SDSS12}, containing several million entries. The angular separation between X-ray and optical positions was required to be smaller than 10\arcsec and the normalised separation (defined as the ratio between the angular separation and the positional error) was lower than or equal to 4. Galaxy clusters were excluded to retain point-like sources only (4XMM catalogue parameter SC$_-$EXTENT < 5\arcsec) and the detection significance was relatively high (4XMM catalogue parameter SC$_-$DET$_-$ML > 14). The resulting cross-matched catalogue has 133,445 X-ray sources.

We restricted our sample to SDSS spectroscopic sources, yielding 21,944 objects, and further required the availability of SDSS spectral fits from the GALSPEC measurements \citep{BrinchmannMNRAS04,KauffmannMNRAS03b,TremontiApJ04} provided by the MPA–JHU DR7 of spectroscopic measurements, resulting in a sample of 2,628 objects. We used seven emission lines (H$\beta$, [\ion{O}{iii}]\,5007, [\ion{O}{i}]\,6300, H$\alpha$, [\ion{N}{ii}]\,6583, [\ion{S}{ii}]\,6716,\,6731), which are commonly used to construct the BPT/VO87 diagrams, to benchmark our synthetic spectral model. The availability of these emission lines further limited the sample size. The emission line fluxes were not corrected for dust attenuation.

Based on a 5-$\sigma$ full width at half maximum (FWHM) threshold of 1000 km s$^{-1}$ for the H$\alpha$ and H$\beta$ Balmer lines \citep{CaccianigaAA08}, our sample was separated into 67 broad emission-line (BEL; type I) galaxies and 1,819 narrow emission-line (NEL; type II) galaxies. We note that the H$\beta$-based FWHM selection restricts our sample to objects with redshift z $\leq$ 0.9.

The AGN sample was further restricted to 751 objects based on their X-ray luminosity, selecting sources with $\mathrm{L_{2-10~\mathrm{keV}}} >$ 10$^{42}$ erg s$^{-1}$ in order to minimise contamination from star-forming galaxies and composite objects. Finally, by requiring the S/N of all seven emission lines to exceed 5, we obtained a final sample consisting of 136 NEL galaxies and 1 BEL galaxy.\

\subsection{Benchmark on the BPT/VO87 diagrams}
\label{sec: benchmark/BPT}

\begin{figure*}[h!]
\centering
\includegraphics[width=18cm]{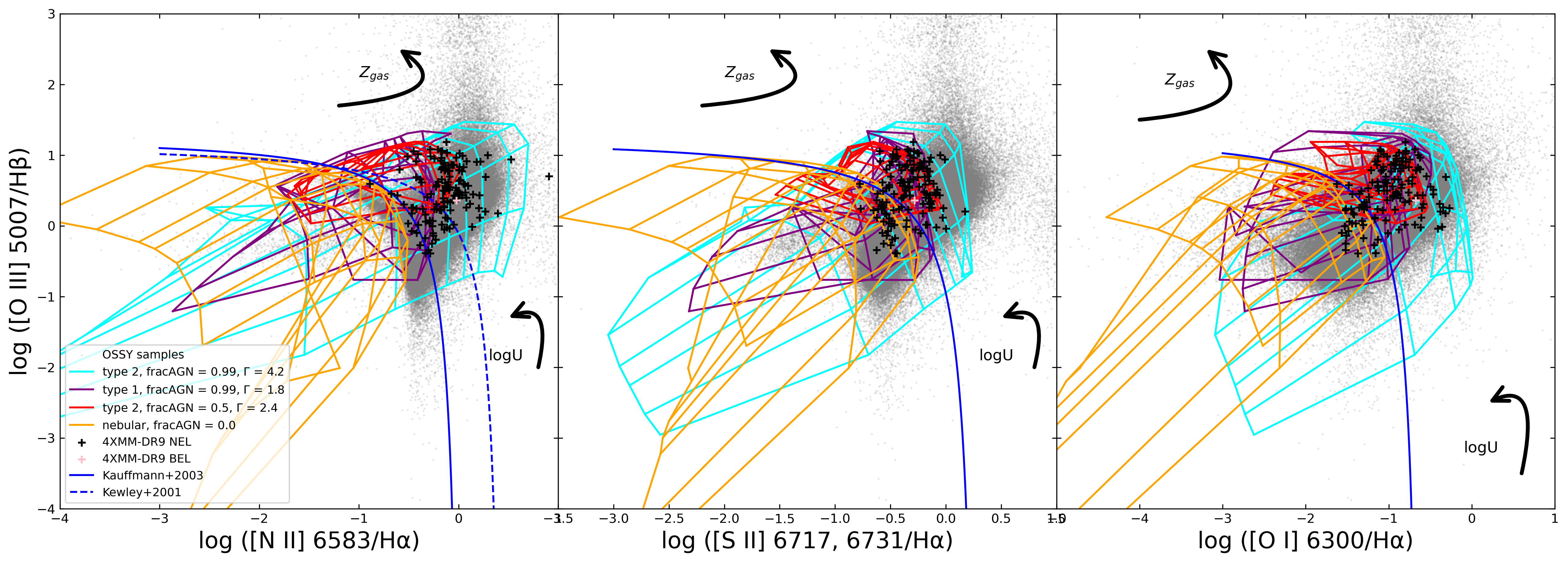}
\caption{Comparison between observations and CIGALE simulations in three commonly used BPT/VO87 diagrams. The OSSY data are represented by grey dots, and the different coloured crosses indicate NEL and BEL X-ray-selected AGNs. The different coloured grids correspond to different CIGALE simulations with different NLR ionization parameters varying from log$\textbf{U}_{\mathrm{NLR}}$ = $-3.8$ to $-1.5$ and metallicities $\mathrm{Z_{gas}}$ ranging from 0.011 to 0.05, with constant densities of log $\mathrm{nH_{NLR}[cm^{-3}]}$ = 3.0 and log $\mathrm{nH_{BLR}[cm^{-3}]}$ = 10.0. Black arrows indicate the directions of increasing log$\textbf{U}$ and $\mathrm{Z_{gas}}$. The demarcation lines separating star-forming galaxies and AGNs as defined by \citet{Kewley2001} (dashed blue line) and \citet{KauffmannMNRAS03a} (solid blue line) are displayed.
}
\label{Fig:BPT}
\end{figure*}

We plot the OSSY sample and the X-ray-selected AGN sample in the BPT/VO87 diagrams and examine whether the simulated emission-line ratios of AGN host galaxies generated by CIGALE with the \texttt{[nebular\_AGN]} module can reproduce these observations. By adopting different parameter settings, we generated four sets of simulated data: \textit{(i)} pure AGNs with \texttt{fracAGN} = 0.99, $\mathrm{\Gamma}$  = 3.0, \textit{(ii)} pure AGNs with \texttt{fracAGN} = 0.99, $\mathrm{\Gamma}$ = 1.8, \textit{(iii)} AGNs mixed with star formation with \texttt{fracAGN} = 0.5, $\mathrm{\Gamma}$ = 2.4 and \textit{(iv)} non-AGN galaxies whose emission lines arise entirely from the stellar population and surrounding \ion{H}{ii} regions (\texttt{fracAGN} = 0.0). Each set of simulations covers a range of ionization parameters \textbf{U} and metallicities $\mathrm{Z_{gas}}$; their output emission-line ratios are plotted in Fig.~\ref{Fig:BPT} as grids consisting of iso-log\textbf{U} and iso-metallicity curves in the BPT/VO87 diagrams, illustrating the range of emission-line ratios covered by our simulations. The model parameters for the four simulations are shown in Table~\ref{Table:BPT}.

From Fig.~\ref{Fig:BPT}, we can see that without introducing \texttt{[nebular\_AGN]} (\texttt{fracAGN} = 0.0), the line ratios simulated by CIGALE can only cover the star-forming sequence of the three BPT/VO87 diagrams. After including the \texttt{[nebular\_AGN]} module, the emission-line ratios predicted by CIGALE can cover nearly all X-ray-selected AGNs and the majority of the OSSY sample. The diagnostic grids with $\Gamma = 3.0$ cover 92.60\%, 83.59\%, and 96.71\% of the SDSS sources falling within the AGN regions in the NH$\alpha$, SH$\alpha$, and OH$\alpha$ plots, respectively. However, we note that different coloured diagnostic grids may partially overlap as the metallicity, density, and ionization parameters are varied. This overlap reflects the degeneracies among these parameters.

We also examined the diagnostic grids with $\Gamma = 4.2$ and found that they cover  96.27\%, 90.71\%, and 97.39\% of the SDSS sources falling within the AGN regions in the three diagrams, respectively. These values represent only a marginal improvement compared to the $\Gamma = 3.0$ case. Therefore, it is not worthwhile to adopt such an extreme $\Gamma$ value at the risk of introducing a physically unrealistic bump discussed in Sect.~\ref{sec: model/Fixing the discontinuity} for such a limited gain. We emphasise that these extreme parameter combinations are primarily used to explore the parameter space. They are not required to reproduce the main distribution of SDSS AGNs, and they may correspond to physically unrealistic SED extrapolations.

It is important to emphasise that the region covered by the diagnostic grids in the line-ratio diagrams depends strongly on how the spectral discontinuity in the 5–50~nm range is treated. As described in Sect.~\ref{sec: model/Fixing the discontinuity}, the ionization of emitting species is highly sensitive to photons in this wavelength interval, while the detailed shape of this part of the AGN continuum, particularly in the EUV and soft X-ray regime, remains uncertain due to the lack of direct observational constraints. The construction of truly reliable diagnostic grids will therefore require more accurate models of the AGN EUV/soft X-ray continuum.

From Fig.~\ref{Fig:BPT}, we can see the variation trends of the line ratios as different parameters change. As $\mathrm{Z_{gas}}$ increases, the abundances of metal species increase, leading to an initial increase in all emission-line ratios. However, since metals act as efficient coolants, when the metallicity becomes sufficiently high, enhanced metal cooling significantly reduces the gas temperature, resulting in a general decline in the line ratios. This effect is particularly pronounced in the low $\Gamma$ case. A higher $\Gamma$ introduces more soft X-ray photons, which ionize species more efficiently.

The dependence of the line ratios on log\textbf{U} is more complex. In the regime of log\textbf{U} $\lesssim -2$, [\ion{O}{iii}]\,5007/H$\beta$ increases with increasing log\textbf{U}, as higher ionization parameters convert more O$^{+}$ into O$^{2+}$. In contrast, the dependence on log\textbf{U} of low-ionization line ratios ([\ion{N}{ii}]\,6583/H$\alpha$, [\ion{S}{ii}]\,6716,6731/H$\alpha$, and [\ion{O}{i}]\,6300/H$\alpha$) depends on the metallicity. For the low-metallicity case, where metal cooling is inefficient and the gas temperature remains high, increasing log\textbf{U} ionizes the low-ionization species responsible for these emission lines into higher ionization states. As a result, these line ratios decrease with increasing log\textbf{U}. For the high-metallicity case, the gas temperature is substantially lower. As log\textbf{U} increases, although low-ionization ions are still ionized to higher states, the accompanying rise in temperature significantly enhances the collisional excitation efficiency. At the same time, compared to the low-metallicity case, the high-metallicity gas contains more metal species that can be ionized. Therefore, increasing the ionization parameter will ionize more neutral metal species, and further enhance the emission-line strengths. These effects compensate for the reduced abundance of low-ionization species due to their conversion into higher ionization states, leading to a slight increase in these line ratios.

We note that at very high log\textbf{U} values (log\textbf{U} $>-2$), all line ratios exhibit a declining trend. As discussed in Section~\ref{sec: benchmark/Dors}, this behaviour arises from the adopted stopping criterion used to stop the simulation at the ionization front. Higher ionization parameters require a larger ionized region to consume the ionizing photons and satisfy the stopping condition, resulting in stronger dust attenuation of emission lines.

The effect of $\mathrm{\alpha_{OX}}$ on the line ratios is broadly similar to that of $\Gamma$. $\delta$ modifies the optical slope of the accretion disk, which mainly affects photons with energies below the ionization potentials of the main emitting species. Therefore, it only weakly affect the line ratios, mainly through minor changes in the gas temperature.

For the small number of OSSY objects with extreme values that cannot be covered by our simulation grids, we attribute this discrepancy primarily to the stopping criterion setting, which limits the ability of very high log\textbf{U} models to produce extreme line ratios due to introducing strong dust attenuation as we discussed above. In fact, these extreme [\ion{O}{iii}]\,5007/H$\beta$ values are typically greater than 1.5, with some reaching nearly 3. Such cases are already extremely rare in SDSS data and are difficult to explain within the framework of standard photoionization models.

The situation for [\ion{O}{i}]\,6300/H$\alpha$ is slightly different. Since [\ion{O}{i}]\,6300 originates from neutral oxygen, it is emitted both at the ionization front and in the PDR immediately adjacent to the ionization front. However, our simulation stops at the ionization front, and we lose a part of the [\ion{O}{i}]\,6300 emission. Nevertheless, it remains unclear to what extent the contribution from the PDR can modify the line ratios.

An alternative explanation is that these emission lines do not arise entirely from AGN photoionization. \citet{Dopita1995} showed that LINER galaxies, narrow-line radio galaxies, and cooling-flow emission regions can be modelled in terms of fast shocks in a relatively gas-poor environment, whereas the narrow-line regions associated with Seyfert 2 and 1.5 galaxies can be interpreted as fast shocks in a gas-rich environment. Using the MAPPINGS III shock and photoionization code, \citet{Allen2008} presented a library of fully radiative shock models. The diagnostic grids constructed from these models can successfully cover the observed data in the LINER and AGN regions of the BPT/VO87 diagrams. \citet{Rich2014, Kewley2019} also discussed the potential confusion caused by the starburst–shock scenario and the starburst–AGN scenario in the mixing sequences of the BPT/VO87 diagrams. These studies further suggest that the emission lines of these objects are not produced exclusively by AGN photoionization.

Considering that the role of shocks in CIGALE is beyond the scope of this work, we plan to incorporate shocks in future updates of the code. On the other hand, given that [\ion{O}{iii}]\,5007 and [\ion{S}{ii}]\,6716,\,6731 can be used to trace AGN outflows \citep{Liu2013,Zakamska2016,Davies2020}, especially [\ion{O}{iii}]\,5007, which is a classic tracer of AGN outflows, we suspect that these extreme line ratios are related to AGN outflows, an effect that is not included in our model.

\subsection{Fitting the X-ray-selected AGNs with and without [\textsl{nebular\_AGN}]}
\label{sec: benchmark/X-ray}

\begin{figure*}[h!]
\centering
\includegraphics[width=18cm]{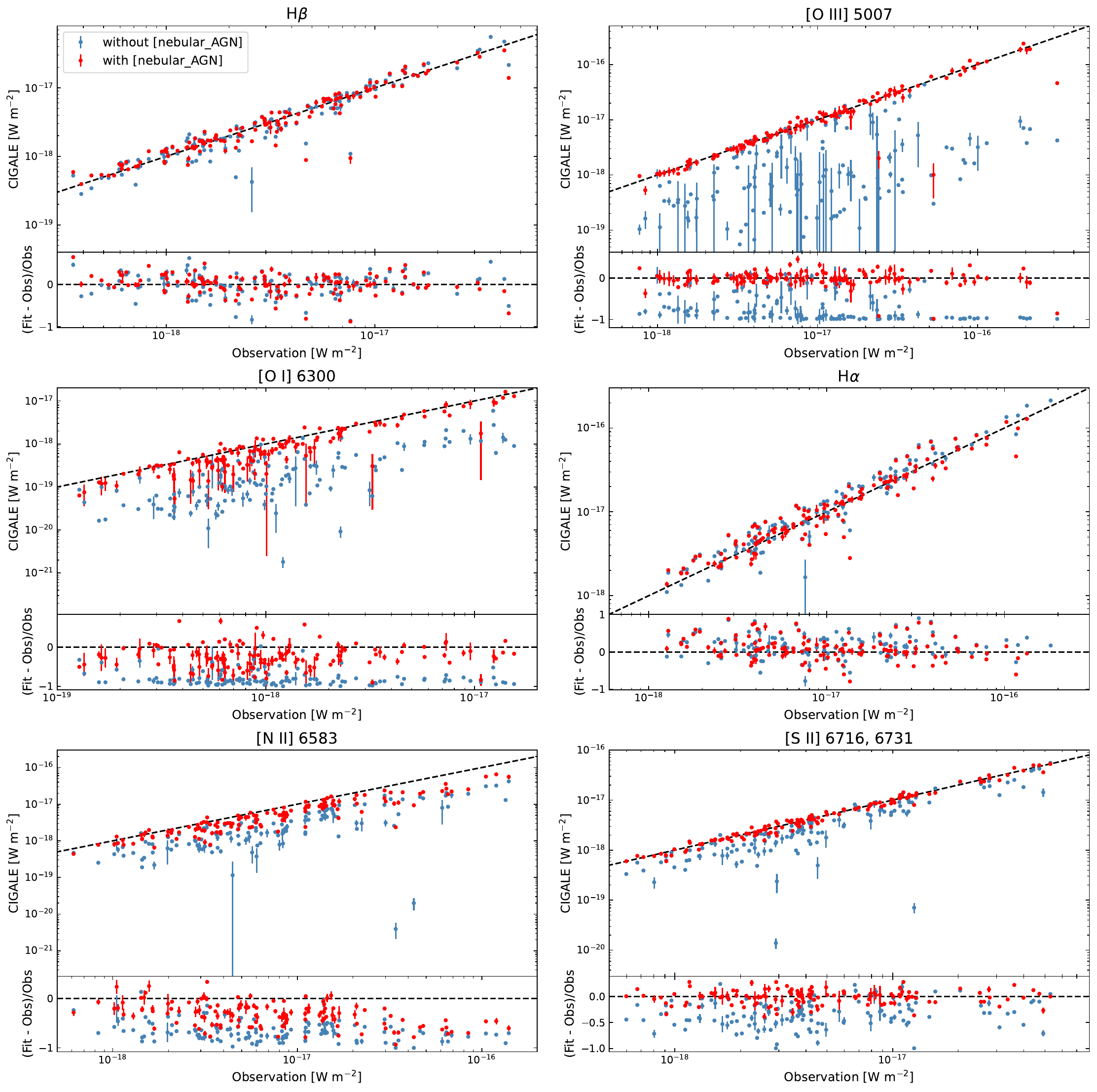}
\caption{Comparison between observed line fluxes and best-fit flux estimates from CIGALE obtained without and with the \texttt{[nebular\_AGN]} module.}
\label{Fig:CIGALE fitting}
\end{figure*}

In this section, we use the X-ray-selected AGN sample consisting of 136 NEL galaxies and 1 BEL galaxy described in Sect.~\ref{sec: benchmark/samples} to verify the improvement in the fitting quality of individual objects by the \texttt{[nebular\_AGN]} module. Specifically, we used as input the photometric data in the five SDSS bands (\textit{u}, \textit{g}, \textit{r}, \textit{i}, \textit{z}) and the fluxes of seven emission lines (H$\beta$, [\ion{O}{iii}]\,5007, [\ion{O}{i}]\,6300, H$\alpha$, [\ion{N}{ii}]\,6583, [\ion{S}{ii}]\,6716,\,6731) for these 137 objects. Using the CIGALE \texttt{pdf\_analysis} mode, we performed fits for these galaxies both with and without the \texttt{[nebular\_AGN]} module. For the case without the \texttt{[nebular\_AGN]} module, the emission lines are entirely produced by the \texttt{[nebular]} module, i.e. they all originate from the excitation of the \ion{H}{ii} regions by the stellar population. Subsequently, we compared the seven best-fit line luminosities estimated by the Bayesian-like algorithm in both cases against the input observed line luminosities, as displayed in Fig.~\ref{Fig:CIGALE fitting}. The parameters used in the two simulations are listed in Table~\ref{Table:fitting check}. 

As Fig.~\ref{Fig:CIGALE fitting} shows, while the H$\mathrm{\alpha}$ and H$\mathrm{\beta}$ Balmer lines are well reproduced without the \texttt{[nebular\_AGN]} module, the module significantly improves the fit of the five metal emission lines. This result is expected because young and hot stellar populations can also produce strong Balmer lines in the surrounding \ion{H}{ii} regions, whereas singly or doubly ionized metal lines often require a harder ionizing radiation field, which is typically associated with AGN photoionization.

Table~\ref{Table:3} shows the changes in the mean relative error (MRE) of the fits to the observed values for five photometric bands and seven emission-line fluxes, with and without the \texttt{[nebular\_AGN]} module. The \texttt{[nebular\_AGN]} module improves the fitting quality of emission-line fluxes without compromising the accuracy of photometric fits, and even significantly improves the $u$-band fit.

\begin{table}[h!]
\caption{Comparison of MREs between fits without and with the \texttt{[nebular\_AGN]} module.}
\tiny
\centering                         
\begin{tabular}{lccc}       
\hline
data & MRE \texttt{[nebular]} & MRE \texttt{[nebular\_AGN]} & $\Delta$MRE\\
\hline
SDSS.$u$ & 0.452 & 0.248 & $-0.204$\\
SDSS.$g$ & 0.140 & 0.112 & $-0.028$\\
SDSS.$r$ & 0.157 & 0.146 & $-0.011$\\
SDSS.$i$ & 0.217 & 0.189 & $-0.028$\\
SDSS.$z$ & 0.235 & 0.199 & $-0.036$\\
\,H$\beta$ & 0.204 & 0.168 & $-0.036$\\
\,[\ion{O}{iii}]\,5007 & 0.728 & 0.113 & $-0.615$\\
\,[\ion{O}{i}]\,6300 & 0.794 & 0.318 & $-0.476$\\
\,H$\alpha$ & 0.267 & 0.232 & $-0.035$\\
\,[\ion{N}{ii}]\,6583 & 0.665 & 0.346 & $-0.319$\\
\,[\ion{S}{ii}]\,6716 & 0.337 & 0.144 & $-0.193$\\
\,[\ion{S}{ii}]\,6731 & 0.365 & 0.172 & $-0.193$\\ 
\hline                                                           
\end{tabular}      
\label{Table:3}     
\end{table}

\section{Discussion}
\label{sec: discussion}

In this section, we compare our diagnostic grids with those from other models to examine their similarities and differences. We further explore the sensitivity of simulated emission-line strengths to the parameters of the \texttt{[nebular\_AGN]} module to identify the most effective line tracers of AGN properties. We also investigate the impact of dust attenuation on the simulated line ratios.

\subsection{Diagnostic grids}
\label{sec: discussion/diagnostic}
Although the \texttt{[nebular\_AGN]} is the first SED-fitting module to incorporate the contribution from the BLR, numerous model grids have already been developed to simulate nebular emission from the NLR \citep[e.g.][]{Groves2004, Feltre2016, Calabro2023, Zhu2023} based on the well-known photoionization synthesis codes \texttt{Cloudy} and \texttt{MAPPINGS}. It is necessary to compare the NLR diagnostic grids implemented in the \texttt{[nebular\_AGN]} module with existing diagnostic grids, in order to assess differences in the adopted photoionization models and their potential impact on the interpretation of observed line ratios.

In this section, we compare our diagnostic grids with the BPT/VO87 diagnostic and UV diagnostic grids presented by \citet{Feltre2016}. This work has been extensively validated and is also based on \texttt{Cloudy}, making it convenient for comparison with our model setup. Moreover, it has served as the foundation for several subsequent developments of emission line diagnostic methods \citep{Hirschmann2023, Mazzolari2024}.

The model setup of \citet{Feltre2016} differs in several aspects from that adopted in the \texttt{[nebular\_AGN]}. First, for the incident radiation field, \citet{Feltre2016} adopted a broken power-law to describe the continuum emission of the accretion disk:
\begin{equation}
   \begin{array}{lll}
     S_\mathrm{{\nu}} &\propto \nu^{\alpha} & ~ 1~\mathrm{nm} \leq\lambda \leq 250~\mathrm{nm} \\
     &\propto \nu^{-0.5} & ~250~\mathrm{nm} <\lambda \leq 10~\mathrm{\mu m} \\
     &\propto \nu^{2} & ~\lambda > 10~\mathrm{\mu m}, \\    
   \end{array}
\label{eq:feltre_field}
\end{equation}
where $-2.0 < \alpha < -1.2$. This model is not fully consistent with the Schartmann model we adopted in the \texttt{[nebular\_AGN]} module. In \citet{Feltre2016}, the disk luminosity $L_{\mathrm{AGN}}$ is fixed at $10^{45}\ \mathrm{erg\ s^{-1}}$. The inner radius of the narrow-line region, $r_{\mathrm{in}}$, is set to 300 pc, corresponding to an incident flux of $L_{\mathrm{AGN}} / 4\pi r_{\mathrm{in}}^{2} \approx 10^{2}\ \mathrm{erg\ s^{-1}\ cm^{-2}}$. When $L_{\mathrm{AGN}}$ for the \citet{Feltre2016} models is scaled to match the observed data, it effectively means that the inner radius is scaled accordingly. This differs from our approach, in which the flux of ionizing photons striking the illuminated face of the gas cloud is directly specified via log\textbf{U}, allowing $L_{\mathrm{AGN}}$ and $r_{\mathrm{in}}$ to be flexibly adjusted according to the input data. In addition, the broken power-law of \citet{Feltre2016} extends down to 1~nm (1.24~keV), without an additional coronal component.

\begin{figure}[h!]
\centering
\includegraphics[width=8cm]{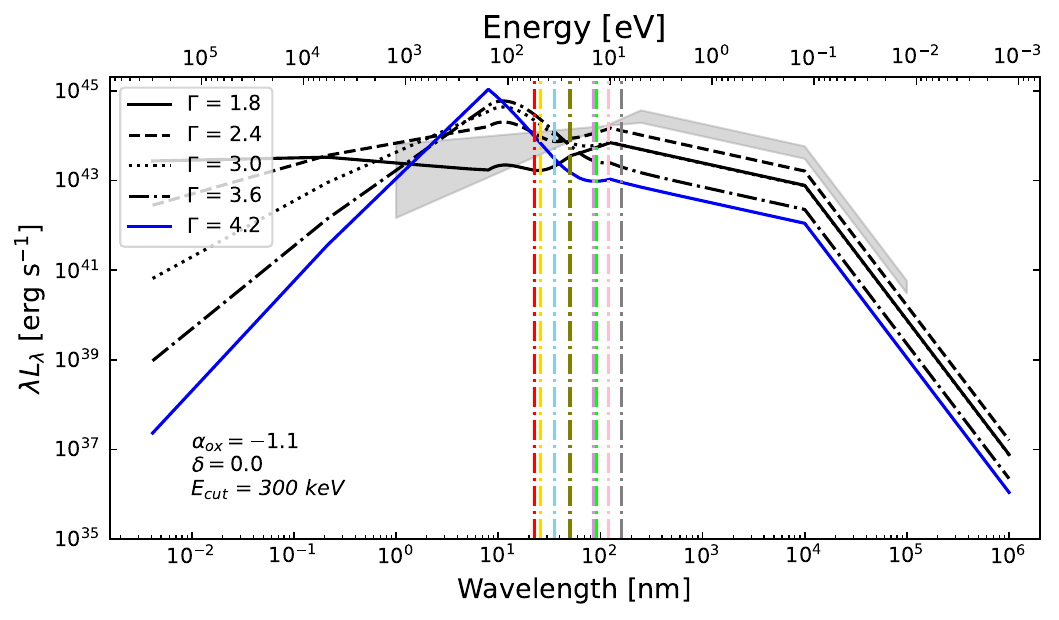}
\caption{Comparison of different incident radiation fields used in our models and in \citet{Feltre2016}. We use lines with different styles to represent the incident radiation fields with different $\Gamma$ values adopted in the \texttt{[nebular\_AGN]} module. Their luminosities are all scaled to the same value as in \citet{Feltre2016}, i.e. $L_{\mathrm{AGN}} = 10^{45}\ \mathrm{erg\ s^{-1}}$. The grey shaded area indicates the accretion disk spectra used as incident radiation fields in the models of \citet{Feltre2016}, with power-law indices between $\alpha = -2.0$ and $-1.2$. The vertical dash-dotted lines, shown in different colours, mark the ionization potentials of a series of atomic and ionic species responsible for AGN emission lines (H: black; C$^{2+}$: gold; C$^{+}$: orange; He: olive; He$^{+}$: red; Mg: grey; O: lime; O$^{+}$: sky-blue; N: violet; S: pink).}
\label{Fig:Feltre incident radiation field}
\end{figure}

 In Fig.~\ref{Fig:Feltre incident radiation field}, we compare the incident radiation fields of our module with those of \citet{Feltre2016}. The incident radiation fields in the \texttt{[nebular\_AGN]} module cover a broader wavelength range at both the short- and long-wavelength ends. For $\lambda \geq 250$ nm, the disk continuum has an identical spectral shape in both cases. At the high-energy end of the spectrum, although the incident radiation fields in \texttt{[nebular\_AGN]} overlap to some extent with the grey shaded region, their shapes are not fully consistent in the 13.6–100~eV range. Given that this part of the spectrum lies very close to the ionization potentials of the main line-emitting species, this inconsistency is one of the potential causes of differences in line ratios between the two grids.

The models of \citet{Feltre2016} adopt hydrogen number densities consistent with those adopted in our setup, i.e. $\log (\mathrm{n_{H}/cm}^{-3}) = 2, 3, 4$. Their metallicity options also closely match those adopted in our models.

The models of \citet{Feltre2016} assume an open geometry, in the sense that scattering from other gas clouds is not taken into account, which differs from our closed geometry. They chose to parameterise their models using the ionization parameter at the Strömgren radius, $\mathbf{U}_{s}$. Given that $L_{\mathrm{AGN}}$, $r_{\mathrm{in}}$, and $\mathrm{n_{H}}$ are already specified, variations in $\mathbf{U}_{s}$ effectively correspond to changes in the volume-filling factor of the gas (i.e. the ratio between the volume-averaged hydrogen density and $\mathrm{n_H}$), as well as, equivalently, changes in the characteristic size of the gas cloud (see Eq.~4 in \citet{Feltre2016}). This differs from our approach, in which we specify the strength of the incident radiation field using the ionization parameter at the inner edge of the narrow-line region, while the gas cloud size is constrained through the stopping criterion.

The models of \citet{Feltre2016} also treat the dust-to-heavy element mass ratio $\xi_{d}$ as a free parameter, with three possible values: 0.1, 0.3, and 0.5. In contrast, our models adopt the default ISM dust settings in \texttt{Cloudy v23.01}, which are scaled self-consistently with metallicity variations.

\begin{figure}[h!]
\centering
\includegraphics[width=8.5cm]{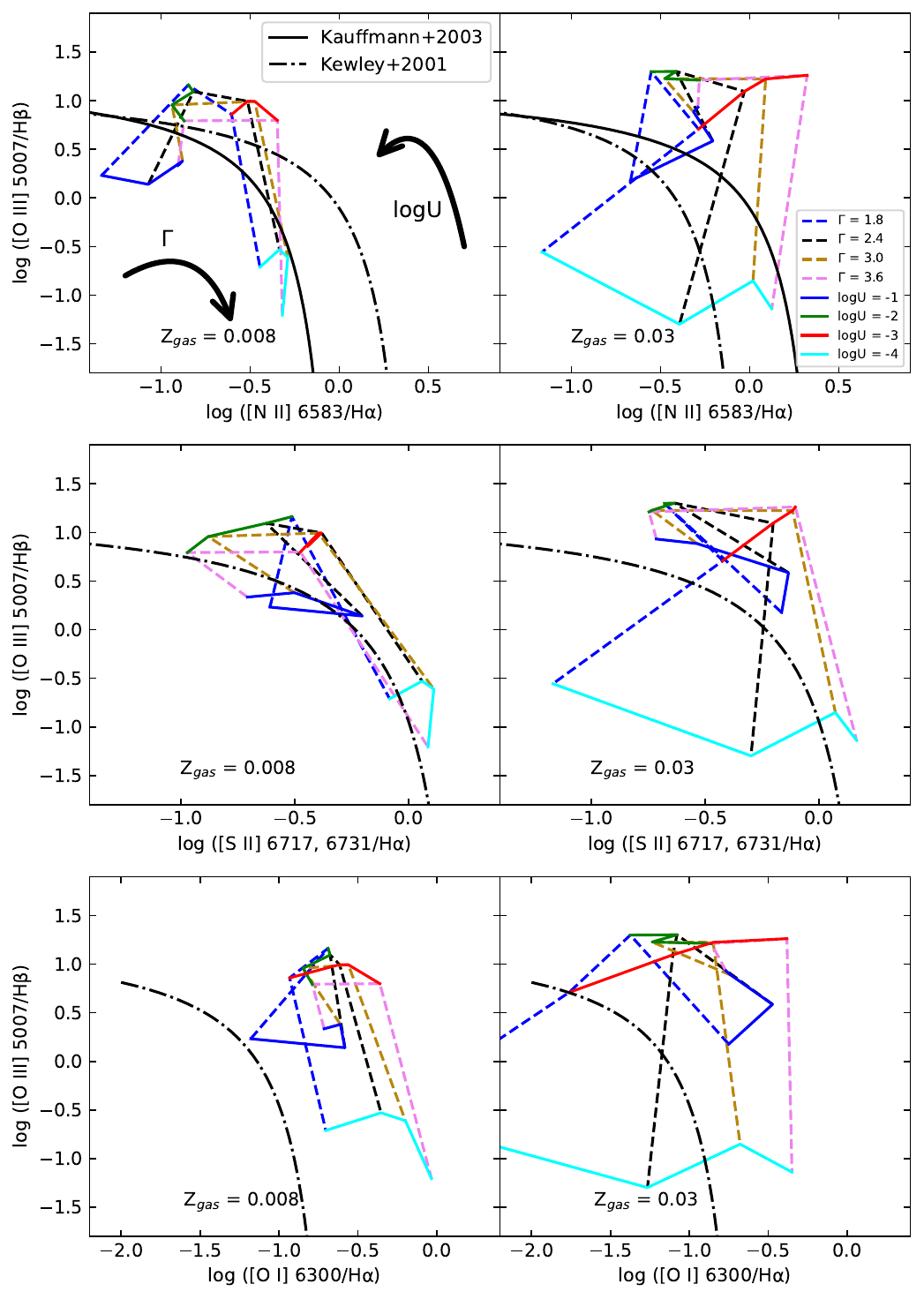}
\caption{BPT/VO87 diagnostic grids constructed using the \texttt{[nebular\_AGN]} module with different ionization parameters and photon indices compared with the diagnostic grids shown in Figs.~2,~3, and 4 of \citet{Feltre2016}.}
\label{Fig:BPT_Feltre}
\end{figure}

In Figs.~2, 3, and 4 of \citet{Feltre2016}, diagnostic grids are presented for $\mathrm{n_{H}} = 10^{3}\,\mathrm{cm^{-3}}$ and gas metallicities $\mathrm{Z_{gas}}$ = 0.008 and 0.03, constructed from combinations of log$\mathbf{U}_{s}=-1, -2, -3, -4 $ and $\alpha=-2.0, -1.7, -1.4, -1.2$. Similarly, in Fig.~\ref{Fig:BPT_Feltre}, we adopted the same hydrogen number density and metallicity values and construct diagnostic grids for our models using log\textbf{U} = -1, -2, -3, -4 and $\Gamma = 1.8, 2.4, 3.0, 3.6$, and the resulting grids are then compared with the grids in Figs.~2, 3, and 4 of \citet{Feltre2016}. We set $\mathrm{\alpha_{OX}}$ to -1.1 for our diagnostic grids.

We find that although our photoionization model setup differs in many aspects from that of \citet{Feltre2016}, the regions covered by the diagnostic grids of the two models in the BPT/VO87 diagrams are broadly similar, indicating that the physical interpretations implied by the two models are consistent. A notable difference is that, for cases with log\textbf{U} = -4, all [\ion{O}{iii}]\,5007/H$\beta$ ratios in our diagnostic grids are lower by a factor of 3--10 compared to those in the cases with log$\mathbf{U}_{s} = -4$ in \citet{Feltre2016}. The most extreme case occurs at $\mathrm{Z_{gas}} = 0.03$, $\Gamma = 1.8$, and log\textbf{U} = -4, where all corresponding line ratios are significantly lower than those of the corresponding model in \citet{Feltre2016}. This discrepancy is likely due to the different definitions of the ionization parameter adopted in the two models. The Str\"omgren ionization parameter $\mathbf{U}_{s}$ is lower than the ionization parameter \textbf{U} at the illuminated face. Therefore, when the \textbf{U} in our model and the $\mathbf{U}_{s}$ in \citet{Feltre2016} are set equal, the model of \citet{Feltre2016} is illuminated by a stronger incident radiation field. This effect becomes more pronounced at lower ionization parameters. In addition, as shown in Fig.~\ref{Fig:Feltre incident radiation field}, the incident radiation field with $\Gamma = 1.8$ provides a significantly smaller flux of ionizing photons in the 10--100 eV range than any of the incident radiation fields used in \citet{Feltre2016}. The combination of low $\Gamma$ and low log\textbf{U} results in a very limited flux of ionizing photons, while the low temperature caused by high metallicity further suppresses the emission line strengths. These combined effects ultimately leading to significantly reduced line ratios in the case with $\mathrm{Z_{gas}} = 0.03$ and $\Gamma = 1.8$.

Another clear difference is that, in our models, the dependence of line ratios on \textbf{U} and $\Gamma$ is not as monotonic as in the models of \citet{Feltre2016}. Regarding log\textbf{U}, as discussed in Sect.~\ref{sec: benchmark/Dors}, our stopping criterion causes models with high log\textbf{U} to experience stronger dust attenuation, which in turn decreases the [\ion{O}{iii}]\,5007/H$\beta$ ratio. Although $\Gamma$ and the spectral index $\alpha$ in \citet{Feltre2016} both control the slope of the X-ray-to-EUV part of the incident radiation field, $\Gamma$ affects a broader wavelength range, covers a wider parameter space, and has a more complex impact on the shape of the incident radiation field. This leads to a non-monotonic dependence of the ionizing photon budget on $\Gamma$ for photons capable of ionizing different species, particularly for species with lower ionization potentials. As shown in Fig.~\ref{Fig:Feltre incident radiation field}, the relative ordering of the ionizing photon fluxes provided by spectra with different photon indices changes across different energy intervals (10--25 eV, 25--50 eV, and 50--500 eV). This complex energy-dependent behaviour, combined with the wavelength dependence of ionization cross-sections of species (as discussed in Sect.~\ref{sec: model/inciden radiation field/corona}), leads to a non-monotonic response of line ratios as a function of $\Gamma$. This is in contrast to the models of \citet{Feltre2016}, where the flux of photons capable of ionizing most emitting species increases monotonically with $\alpha$.

\begin{figure}[h!]
\centering
\includegraphics[width=6cm]{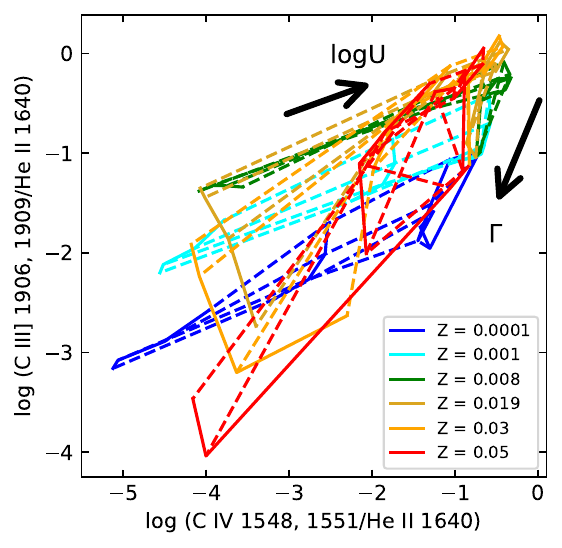}
\caption{Ultraviolet diagnostic grids built using the \texttt{[nebular\_AGN]} module, with $\mathrm{n_{H}}$ = 10$^{3}$ cm$^{-3}$, log\textbf{U} = -1, -2, -3, -4, $\Gamma = 1.8, 2.4, 3.0, 3.6$, and $\mathrm{\alpha_{OX}}$ = -1.1, compared with the diagnostic grids in Fig.~6 of \citet{Feltre2016}.}
\label{Fig:UV_Feltre}
\end{figure}

\citet{Feltre2016} showed that the UV diagnostic grids based on the line ratios \ion{C}{iii}]\,1907,1909/\ion{He}{ii}\,1640 versus \ion{C}{iv}\,1548,1551/\ion{He}{ii}\,1640 can effectively distinguish AGNs from star-forming regions, and presented such grids for different metallicities in Fig.~6 of their work. Similar to the optical line ratio diagnostics, we construct the corresponding UV diagnostic grids using the \texttt{[nebular\_AGN]} module, as shown in Fig.~\ref{Fig:UV_Feltre}.

We find that the UV grids produced by the \texttt{[nebular\_AGN]} module reproduce the overall trends of line ratios with varying physical parameters seen in Fig.~6 of \citet{Feltre2016}, and the regions covered by the two sets of grids are broadly consistent when both line ratios are greater than $-2$. However, the UV grids from the \texttt{[nebular\_AGN]} extend to a much broader region in the high-$\Gamma$ regime, with both line ratios reaching values below $-2$, down to approximately $-4$. This difference arises because the range of $\Gamma$ explored in our models induces a much larger variation in the shape of the incident radiation field from the X-ray to UV regime than the variation produced by the range of power-law indices $\alpha$ in \citet{Feltre2016}. Since \ion{He}{ii}\,1640 is primarily produced by recombination of He$^{2+}$, its strength is directly linked to the ionization of He$^{+}$, which has an ionization potential of 54.4 eV, higher than the ionization potentials required to produce C$^{+}$ and C$^{2+}$ (24.4 eV and 47.9 eV, respectively). In this photon energy range, the flux of ionizing photons generally increases monotonically with $\Gamma$, with the enhancement becoming stronger at higher energies. As a result, the intensity of \ion{He}{ii}\,1640 line increases significantly, driving both line ratios to very low values. For the same reason, our UV diagnostic grids reach slightly lower maximum values on both axes compared to those in \citet{Feltre2016}.

In this section, as well as in Sects.~\ref{sec: benchmark/Dors} and ~\ref{sec: benchmark/BPT}, we show that the model setup and the adopted stopping criteria can affect the predicted line ratios for certain parameter combinations, particularly at high log\textbf{U}. These effects may further influence the inference of physical parameters based on the module outputs. Therefore, in Appendix~\ref{app: reliability}, we assess the reliability of the main physical parameters derived from the current model to help users identify more reliable parameter ranges when using the \texttt{[nebular\_AGN]} module.

\subsection{Line sensitivity}
\label{sec: discussion/sensitivity}

Using emission lines to trace and diagnose the existence and nature of AGNs has been a hot topic in extragalactic astronomy and has undergone a long period of development. The most widely used tracers are the optical emission lines commonly used to construct the BPT/VO87 diagrams: [\ion{O}{iii}]\,5007, [\ion{O}{i}]\,6300, [\ion{N}{ii}]\,6583, and [\ion{S}{ii}]\,6716,\,6731. In recent years, with the launch of JWST, an increasing number of studies have found that high-redshift galaxies generally have lower metallicities and higher ionization parameters \citep{Curti2024, Tacchella2023, Trump2023}, which makes it difficult to robustly identify AGNs using optical emission lines alone. A growing number of studies have shown that high-redshift AGNs detected with JWST overlap with the local star-forming (SF) sequence in the BPT/VO87 diagrams \citep{Harikane2023,Kocevski2023,Maiolino24,Ubler2023}. Therefore, the potential of UV emission lines as AGN tracers has been increasingly investigated, especially in the high-redshift Universe. \citet{Feltre2016} proposed that the luminosity ratios of \ion{C}{iv}\,1550, \ion{O}{iii}]\,1663, \ion{N}{iii}]\,1750, \ion{Si}{iii}]\,1888, and \ion{C}{iii}]\,1908 relative to \ion{He}{ii}\,1640 can serve as effective diagnostics for distinguishing between nuclear activity and star formation. In addition, other line ratios, such as \ion{C}{iv}\,1550/\ion{C}{iii}]\,1908, \ion{N}{v}\,1240/\ion{He}{ii}\,1640 and \ion{N}{v}\,1240/\ion{C}{iv}\,1550, although unable to clearly distinguish AGNs from SF galaxies, are valuable for probing the physical conditions of the ionized gas, such as the ionization parameter and metallicity. Meanwhile, several neon emission lines, such as [\ion{Ne}{iv}]\,2424, [\ion{Ne}{iii}]\,3343, and [\ion{Ne}{v}]\,3426, can also effectively distinguish AGNs and SF galaxies. Owing to their high ionization potentials and relative insensitivity to stellar photoionization, these UV lines are increasingly employed as AGN diagnostics in high-redshift galaxies \citep{Hirschmann2023,Scholtz2025,Treiber2025,Tang_2025}. At longer wavelengths, some mid-IR and far-IR emission lines, such as [\ion{Mg}{v}]\,5.6\,$\mu$m, [\ion{Ne}{v}]\,14.3\,$\mu$m, [\ion{Ne}{v}]\,24\,$\mu$m, and [\ion{O}{iv}]\,25.89\,$\mu$m, are widely used to trace AGN activity, while [\ion{C}{ii}]\,158\,$\mu$m is commonly used to constrain the physical properties of AGN host galaxies \citep{Melendez2008,Stacey2010,Spinoglio1992,Spinoglio_2017}.

In this section, we explore the sensitivity of these emission lines to the parameters of the \texttt{[nebular\_AGN]} module, thereby assessing their potential as indicators of AGN properties. We investigated nine AGN-related parameters in CIGALE (as shown in Table~\ref{Table:parameters}): \texttt{fracAGN}, $\mathrm{\alpha_{OX}}$, $\mathrm{\Gamma}$, $\mathrm{\delta}$, $\mathrm{logU_{NLR}}$, $\mathrm{logU_{BLR}}$, $\mathrm{Z_{gas}}$, log$\,\mathrm{nH_{NLR}}$, and log$\,\mathrm{nH_{BLR}}$. Using the \texttt{savefluxes} mode of CIGALE, we adopted multiple values for each parameter, which are summarised in Table~\ref{Table:line_sensivity}, in order to cover their full parameter ranges. We thereby generated a large number of simulated emission lines corresponding to different combinations of parameter values, covering a wide range of AGN properties. When examining the sensitivity of a specific parameter (the target parameter) to emission lines, we fixed the remaining eight parameters (the remaining parameters) at constant values. This yielded a set of models in which the target parameter varied while the remaining parameters were held fixed. For this set of models, we computed the mean and standard deviation of the logarithmic flux ratios of each emission line relative to H$\alpha$. We then repeated this process for all possible combinations of the remaining parameter values. For each emission line, this procedure produced a set of mean relative fluxes and a set of standard deviations corresponding to all possible combinations of the remaining parameters. From these results, we calculated the mean logarithmic relative fluxes $\overline{\log(\mathrm{line}/\mathrm{H}\alpha)}$ and the average standard deviation of the logarithmic relative fluxes $\overline{\sigma[\log(\mathrm{line}/\mathrm{H}\alpha)]}$, for each emission line.

A good tracer of a given parameter should have sufficient strength to be detected and exhibit significant variation with that parameter. Emission lines with mean logarithmic line ratios $\overline{\log(\mathrm{line}/\mathrm{H}\alpha)} > -2$ are considered bright enough to be potential tracers; those with $\overline{\sigma[\log(\mathrm{line}/\mathrm{H}\alpha)]} > 0.5$ are classified as strongly sensitive to the parameter, while those with $0.3 < \overline{\sigma[\log(\mathrm{line}/\mathrm{H}\alpha)]} < 0.5$ are regarded as weakly sensitive. A summary of common AGN emission lines and their sensitivities to the nine parameters is presented in Table~\ref{Table:tracers}.

In Table~\ref{Table:tracers}, we see that although some emission lines can serve as relatively good tracers, the most commonly used emission lines for diagnosing AGNs are sensitive to multiple parameters. It is expected that no emission line is sensitive to $\mathrm{\delta}$. The parameter $\mathrm{\delta}$ determines the shape of the incident radiation field between 125~nm and 10~$\mu$m and therefore does not significantly change the flux of ionizing photons in the soft X-ray to EUV range. We note that the selection of parameter ranges and model configurations also affects the results in this table. Different choices of parameters and evaluation methods may lead to different conclusions. Therefore, the results in this table should only be regarded as a preliminary reference. A detailed exploration of the underlying physical interpretation is beyond the scope of this work, and we encourage users to investigate the sensitivity of the emission lines to different parameters based on this framework according to their own needs.

\subsection{Dust attenuation}
\label{sec: discussion/attenuation}

\begin{figure}[h!]
\centering
\includegraphics[width=9cm]{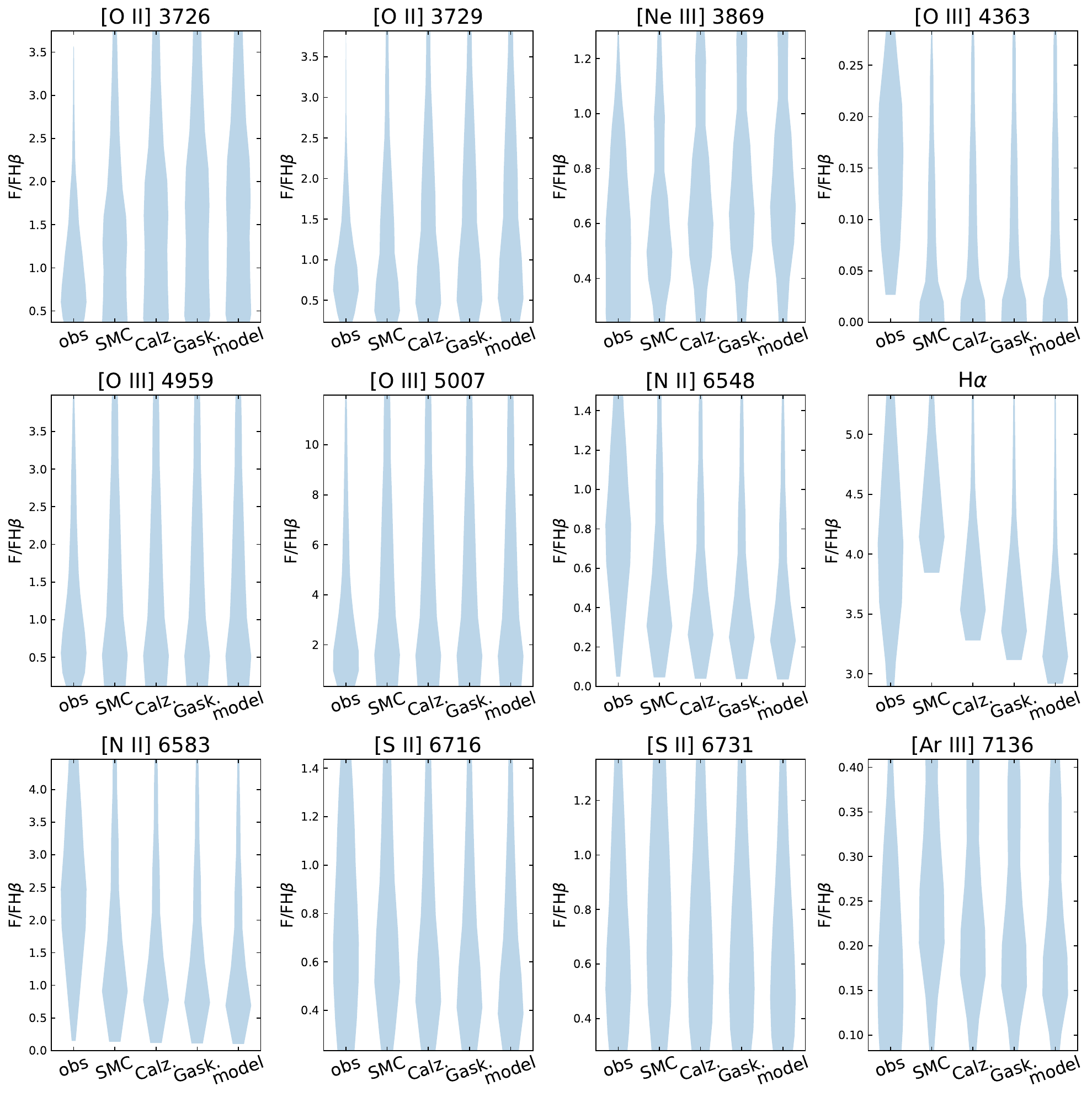}
\caption{Comparison of the distributions of observed NLR emission-line fluxes from the X-ray-selected sample (obs) with simulated emission lines without reddening (model) and with different dust attenuation laws (SMC, Calzetti2000, Gaskell2004) adopting E(B-V) colour excess values of 0.30, 0.10, and 0.05.}
\label{Fig:violin}
\end{figure}

Finally, we discuss the impact of dust attenuation settings on the simulations. As mentioned in Sect.~\ref{sec: model/geometry}, we use the polar dust settings in \texttt{[skirtor2016]} to attenuate both BLR and NLR emission lines. We assume that polar dust is distributed isotropically around the disk, and three different extinction laws are provided in \texttt{[skirtor2016]} for the polar dust: Small Magellanic Cloud (SMC) \citep{Pei1992}, Calzetti2000 \citep{CalzettiApJ00}, and Gaskell2004 \citep{Gaskell2004}. The level of dust extinction is set by a free parameter in \texttt{[skirtor2016]}: the colour excess $\mathrm{E(B-V)}$. To visually demonstrate the effect of dust, we applied three dust attenuation laws with different $\mathrm{E(B-V)}$ values (0.05, 0.1, 0.3) to generate a series of simulated Type II AGNs with varying physical parameters in CIGALE, as shown in Table~\ref{Table:line_dust}. We then compared the distributions of the simulated emission-line fluxes with those of the X-ray-selected NEL AGN sample described in Sect.~\ref{sec: benchmark/samples/X-ray selected samples}, as illustrated in Fig.~\ref{Fig:violin}.

We can see in Fig.~\ref{Fig:violin} that, apart from a few emission lines near H$\beta$, different colour excess values significantly affect the distributions of simulated emission-line ratios. Although different colour excess values do not change the shapes of the distributions, they shift their overall positions. This clearly shows the importance of dust attenuation treatment in SED fitting \citep{MountrichasAA21}. 

\section{Conclusion}
\label{sec: conclusion}
In this work, based on the AGN unified model, as well as the coronal emission model (power-law) and the accretion disk emission model (Schartmann) both implemented in CIGALE, we developed a new \texttt{[nebular\_AGN]} module for the SED fitting code CIGALE using the spectral synthesis code \texttt{Cloudy v23.01}. This module enables the fitting of the emission lines from the BLR and NLR of AGNs over a spectral range from the UV to the far-IR.

We benchmarked the model using multiple approaches. First, by fitting the band and line fluxes of the composite quasar spectrum from \citet{VandenBerkAJ01}, we verified that the \texttt{[nebular\_AGN]} module can approximately reproduce the fluxes of most emission lines associated with AGNs. For the emission lines that are not well reproduced, the discrepancies arise partly from the limitations of our model and partly from the stacking effects inherent in the composite spectrum. We compared the ability of the \texttt{[nebular\_AGN]} module to infer metallicity with the empirical calibration of \citet{Dors2021} based on the direct method, and we found that for specific combinations of parameters, the metallicities derived with the \texttt{[nebular\_AGN]} are consistent with those given by the empirical formula. The differences between the two methods mainly arise from the parameter settings adopted in our model as well as the insufficient consideration of the effects of physical parameters and the hardness of the ionizing radiation field on the empirical method. We then benchmarked the \texttt{[nebular\_AGN]} module against 137 X-ray-selected AGNs and 664,187 galaxies from SDSS DR7 using three commonly used BPT/VO87 diagrams: [\ion{N}{ii}]\,6583/H$\alpha$ vs [\ion{O}{iii}]\,5007/H$\beta$, [\ion{S}{ii}]\,6716,\,6731/H$\alpha$ vs [\ion{O}{iii}]\,5007/H$\beta$, and [\ion{O}{i}]\,6300/H$\alpha$ vs [\ion{O}{iii}]\,5007/H$\beta$. The simulated line ratios cover almost all X-ray-selected AGNs and the majority of SDSS galaxies, further demonstrating the reliability of the \texttt{[nebular\_AGN]} module. Only a small number of SDSS galaxies with extreme emission-line ratios are not covered by the simulations. This is partly due to our choice of stopping criterion at the ionization front and partly due to the absence of PDRs, shocks, or outflows in our model. If outflows contribute significantly to AGN emission lines, our module may tend to overestimate $\mathrm{\alpha_{OX}}$ and $\mathrm{\Gamma}$. Finally, we applied the \texttt{[nebular\_AGN]} module to directly fit the band fluxes and main emission line fluxes of our 137 X-ray-selected AGNs, and we found that it significantly improves the fitting quality of metal-line fluxes.

We compared our photoionization model diagnostic grids with the widely used diagnostic grids of \citet{Feltre2016} and discussed the origins of their similarities and differences. We analysed the sensitivity of emission lines to the different model parameters to provide a preliminary assessment of which emission lines are effective at tracing the physical parameters in our models. We identified a set of emission lines that are sensitive to the different model parameters and therefore have the potential to be used as diagnostics of AGN properties. Additionally, we demonstrated that the dust attenuation law and the choice of colour excess play a crucial role in the simulations. 

Future improvements to our model will focus on incorporating contributions from the PDRs, shocks, and outflows to better reproduce both low- and high-ionization lines. Spatially resolved observations of the ionization structure of nearby AGNs will help us adopt more physically motivated stopping criteria. More sophisticated kinematic models could also be employed to generate line profiles that better match observations. In addition, this work shows that the quality of the simulations depends critically on the accuracy of the soft X-ray model of the warm corona, particularly on the spectral shape in the 13.6--100~eV energy range, where the emission is heavily absorbed and therefore cannot be directly constrained by observations. Joint analyses combining X-ray, UV, and optical observations with photometric data and emission-line ratios may provide a promising way to constrain the spectral shape in this energy range, thereby offering valuable constraints on AGN corona models. Another important step is to couple AGN emission with host galaxy emission, including the attenuation of emission by gas and dust.

The incorporation of the AGN emission lines in the framework of the SED fitting code CIGALE is essential in understanding the co-evolution of black holes and their host galaxies in the high-redshift Universe with present or future facilities, such as JWST, PFS, MOONS, or PRIMA \citep{BisigelloAA24}. Estimating black hole parameters from observables can enable discrimination between different models of black hole formation, growth, and merging \citep{Volonteri25}.

\begin{acknowledgements}
This work was supported by the Thematic Actions 'Physique et Chimie du Milieu Interstellaire' (PCMI) of  and 'Cosmologie et Galaxies' (CG) from INSU Programme National 'Astro', with contributions from CNRS Physique $\mathrm{\&}$ CNRS Chimie, IN2P3, CEA, and CNES. M. Boquien acknowledges support by the ANID BASAL project FB210003. This work was supported by the French government through the France 2030 investment plan managed by the National Research Agency (ANR), as part of the Initiative of Excellence of Université Côte d’Azur under reference No. ANR-15-IDEX-01.
\end{acknowledgements}

\bibliographystyle{aa} 
\bibliography{references_AGN}

\begin{appendix}
\clearpage

\section{Parameter list}
\label{app: parameter_lis}
\noindent
\begin{minipage}{\textwidth}
\captionof{table}{List of parameters in the \texttt{[nebular\_AGN]} and related parameters from other modules.}
\centering    
\begin{tabular}{|l|l|p{4cm}|p{4cm}|}
\hline
\multirow{2}{*}{\texttt{[yang20]}} 
& gam ($\mathrm{\Gamma}$) &\textbf{1.8}, 2.4, 3.0, 3.6, 4.2, 4.8 & AGN photon index\\
& & & \\
& alpha\_ox ($\mathrm{\alpha_{OX}}$) & -1.9, -1.5, \textbf{-1.1} & UV-to-X-ray spectral slope\\
\hline
\multirow{2}{*}{\texttt{[skirtor2016]}} 
& fracAGN & [0 - 1) & Fraction of the total IR luminosity contributed by the AGN\\
& & & \\
& i & 0, 10, 20, \textbf{30}, 40, 50, 60, 70, 80, 90 & Viewing angle \\
& & & \\
& oa & 10, 20, 30, \textbf{40}, 50, 60, 70, 80 & Angle between the equatorial plane and edge of the torus\\
& & & \\
& delta ($\mathrm{\delta}$) & -0.5, \textbf{0.0}, 0.5 & Power-law index modifying the optical slope of the disk \\
\hline
\multirow{2}{*}{\texttt{[nebular\_AGN]}} 
& logU\_NLR & [-1, -4] by step size 0.1 & Ionization parameter on the illuminated face of the NLR\\
& & & \\
& logU\_BLR & [-1, -4] by step size 0.1 & Ionization parameter on the illuminated face of the BLR\\
& & & \\
& nH\_NLR (log$\,\mathrm{nH_{NLR}}$) & 2, \textbf{3}, 4 & Hydrogen density of the NLR\\
& & & \\
& nH\_BLR (log$\,\mathrm{nH_{BLR}}$) & 8, \textbf{10}, 12 & Hydrogen density of the BLR\\
& & & \\
& metallicity ($\mathrm{Z_{gas}}$) & [0.00001, ..., \textbf{0.014}, ..., 0.05] & Gas metallicity of the BLR and the NLR\\
& & & \\
& f\_NLR & [0 - 1] & Covering factor of the NLR\\
& & & \\
& f\_BLR  & [0 - 1] & Covering factor of the BLR\\
& & & \\
& lines\_width\_NLR & \textbf{300} (free parameter) & Emission-line width of the NLR\\
& & & \\
& lines\_width\_BLR & \textbf{800} (free parameter) & Emission-line width of the BLR\\
& & & \\
& agn\_emission & \textbf{True}/False & Boolean parameter controlling whether emission lines are included in the simulation\\
\hline
\end{tabular}
\tablefoot{The default values are in bold. \texttt{agn\_emission} is used to determine whether to include emission lines in the simulation.}
\label{Table:parameters}     
\end{minipage}

\clearpage

\section{Table of input model parameters}
\label{app: mp}
\noindent
\begin{minipage}{\textwidth}
\captionof{table}{Model parameters and their adopted values for the four simulations presented in Sect.~\ref{sec: benchmark/BPT}.}
\centering
\begin{tabular}{llp{3cm}p{3cm}p{3cm}p{3cm}}
\hline
module & parameter & fracAGN=0.0 & fracAGN=0.5 & fracAGN=0.99 & fracAGN=0.99 \\
\hline
\multirow{9}{*}{\textbf{[nebular]}} 
& logU & $-3.8, -3.5, -3.0,$ $-2.7, -2.5, -2.0,$ $-1.3$ & $-2.0$ & $-2.0$ & $-2.0$\\
\\
& zgas & $0.0001, 0.001, 0.003,$ $0.006, 0.011, 0.020,$ $0.041$ & $0.011$ & $0.02$ & $0.02$\\
\\
&  ne &100 & 100 & 100 & 100\\
\hline
\multirow{2}{*}{\textbf{[skirtor2016]}} 
& oa & - & 40 & 40 & 40\\
& i & - & 60 & 60 & 60\\
\hline
\multirow{2}{*}{\textbf{[yang20]}} 
& gam & - & 2.4 & 1.8 & 3.0\\
& alpha\_ox & - & $-1.1$ & $-1.1$ & $-1.1$\\
\hline
\multirow{10}{*}{\textbf{[nebular\_AGN]}} 
& metallicity & - & $0.00001, 0.001, 0.005,$ $0.011, 0.014, 0.019,$ $0.033, 0.05$ & $0.00001, 0.001, 0.005,$ $0.011, 0.014, 0.019,$ $0.033, 0.05$ & $0.00001, 0.001, 0.005,$ $0.011, 0.014, 0.019,$ $0.033, 0.05$\\
\\
& logU\_NLR & - & $-3.8, -3.5, -3.2, -3.0,$ $-2.5, -2.0, -1.5$  & $-3.8, -3.5, -3.2, -3.0,$ $-2.5, -2.0, -1.5$ & $-3.8, -3.5, -3.2, -3.0,$ $-2.5, -2.0, -1.5$ \\
\\
& logU\_BLR & - & $-3.8, -3.5, -3.2, -3.0,$ $-2.5, -2.0, -1.5$  & $-3.8, -3.5, -3.2, -3.0,$ $-2.5, -2.0, -1.5$ & $-3.8, -3.5, -3.2, -3.0,$ $-2.5, -2.0, -1.5$ \\
\\
& f\_NLR & - & 0.2 & 0.2 & 0.2 \\
\\
& f\_BLR & - & 0.2 & 0.2 & 0.2 \\
\hline
\end{tabular}
\tablefoot{By varying \texttt{fracAGN}, we generate simulated AGN emission lines with different AGN contributions. Since the BPT/VO87 diagrams are narrow-line diagnostics, all simulated AGNs are configured as type II sources through appropriate combinations of \texttt{oa} and \texttt{i}. The modules used in the simulations are: \texttt{[sfhdelayed]}, \texttt{[bc03]}, \texttt{[nebular]}, \texttt{[dustatt\_modified\_starburst]}, \texttt{[dl2014]}, \texttt{[skirtor2016]}, \texttt{[yang20]}, \texttt{[nebular\_AGN]}, and \texttt{[redshifting]}. Parameters not shown in the table are set to their default values.}
\label{Table:BPT}     
\end{minipage}

\vspace{0.5cm}

\noindent
\begin{minipage}{\textwidth}
\captionof{table}{Model parameters and their adopted values for the two fitting runs presented in Sect.~\ref{sec: benchmark/X-ray}.}
\centering
\begin{tabular}{llp{5cm}}
\hline
module & parameter & value\\
\hline
\multirow{2}{*}{\textbf{[sfhdelayed]}} 
& tau\_main & 500, 1000, 5000\\
& age\_main & 500, 1000, 3000\\
\hline
\multirow{1}{*}{\textbf{[bc03]}} 
& metallicity & 0.0004, 0.008, 0.05\\
\hline
\multirow{2}{*}{\textbf{[nebular]}} 
& logU & $-3.5$, $-2.5$, $-1.5$\\
& zgas & 0.0001, 0.002, 0.014, 0.041\\
\hline
\multirow{1}{*}{\textbf{[dustatt\_modified\_starburst]}} 
& E\_BV\_factor & 0.03, 0.2, 0.44\\
\hline
\multirow{3}{*}{\textbf{[skirtor2016]}} 
& oa & 40 \\
& i & 70 (type II)/30 (type I)\\
& fracAGN & 0.1, 0.3, 0.5, 0.7, 0.99\\
\hline
\multirow{2}{*}{\textbf{[yang20]}}
& gam & 1.8, 3.0\\
& alpha\_ox & -1.1\\
\hline
\multirow{5}{*}{\textbf{[nebular\_AGN]}}
& metallicity & 0.001, 0.006, 0.014, 0.025, 0.05\\
& f\_NLR & 0.1, 0.2, 0.3 \\
& f\_BLR & 0.2 (type II), 0.1, 0.2, 0.3 (type I)\\
& logU\_NLR & $-4.0$, $-3.0$, $-2.0$, $-1.0$\\
& logU\_BLR & $-4.0$, $-3.0$, $-2.0$, $-1.0$\\
\hline
\end{tabular}
\tablefoot{By varying the opening angle \texttt{oa}, the viewing angle \texttt{i}, and \texttt{fracAGN}, we allow the fitting procedure to explore both type I and type II AGNs with different AGN contributions. The modules used in the fitting are: \texttt{[sfhdelayed]}, \texttt{[bc03]}, \texttt{[nebular]}, \texttt{[dustatt\_modified\_starburst]}, \texttt{[dl2014]}, \texttt{[skirtor2016]}, \texttt{[yang20]}, \texttt{[nebular\_AGN]}, and \texttt{[redshifting]}. Parameters not shown in the table are set to their default values.}                 
\label{Table:fitting check}     
\end{minipage}
\clearpage

\noindent
\begin{minipage}{\textwidth}
\captionof{table}{Model parameters and their adopted values for the emission-line sensitivity analysis presented in Sect.~\ref{sec: discussion/sensitivity}.}
\centering
\begin{tabular}{llp{6cm}}
\hline
\multirow{4}{*}{\textbf{[skirtor2016]}} 
& i & 0\\
& delta & $-0.5$, 0.0, 0.5 \\
& fracAGN & 0.0, 0.1, 0.2, 0.3, 0.4, 0.5, 0.6, 0.7, 0.8, 0.9, 0.99\\
\hline
\multirow{2}{*}{\textbf{[yang20]}}
& gam & 1.8, 2.4, 3.0, 4.2 \\
& alpha\_ox & $-1.9, -1.5, -1.1$\\
\hline
\multirow{9}{*}{\textbf{[nebular\_AGN]}}
& metallicity & 0.00001, 0.001, 0.004, 0.006, 0.014, 0.019, 0.025, 0.041\\
& nH\_NLR & 2.0, 3.0, 4.0 \\
& nH\_BLR & 8.0, 10.0, 12.0 \\
& f\_NLR & 0.2 \\
& f\_BLR & 0.2\\
& logU\_NLR & $-3.8, -3.5, -3.3, -3.0, -2.7, -2.5, -2.3, -2.0,$ $-1.5, -1.3$\\
& logU\_BLR & $-3.8, -3.5, -3.3, -3.0, -2.7, -2.5, -2.3, -2.0,$ $-1.5, -1.3$\\
\hline
\end{tabular}
\tablefoot{The modules used in the simulations are: \texttt{[sfhdelayed]}, \texttt{[bc03]}, \texttt{[nebular]}, \texttt{[dustatt\_modified\_starburst]}, \texttt{[dl2014]}, \texttt{[skirtor2016]}, \texttt{[yang20]}, \texttt{[nebular\_AGN]}, and \texttt{[redshifting]}. Parameters not shown in the table are set to their default values.}
\label{Table:line_sensivity}     
\end{minipage}

\vspace{1cm}

\noindent
\begin{minipage}{\textwidth}
\captionof{table}{Model parameters and their adopted values used to assess the effect of dust attenuation on emission lines in Sect.~\ref{sec: discussion/attenuation}.}
\centering
\begin{tabular}{llp{6cm}}
\hline
\multirow{3}{*}{\textbf{[skirtor2016]}} 
& i & 60\\
& delta & $-0.5$, $0.0$, $0.5$ \\
& fracAGN & 0.99\\
\hline
\multirow{2}{*}{\textbf{[yang20]}}
& gam & 1.8, 2.4, 3.0, 3.6, 4.2 \\
& alpha\_ox & $-1.1$\\
\hline
\multirow{9}{*}{\textbf{[nebular\_AGN]}}
& metallicity & 0.004, 0.005, 0.006, 0.011, 0.014, 0.019, 0.033\\
& nH\_NLR & 2.0, 3.0, 4.0 \\
& nH\_BLR & 8.0, 10.0, 12.0 \\
& f\_NLR & 0.2 \\
& f\_BLR & 0.2\\
& logU\_NLR & $-3.8, -3.7, -3.6, -3.5, -3.4, -3.3, -3.2, -3.1,$ $-3.0, -2.8, -2.5, -2.3, -2.0, -1.5$\\
& logU\_BLR & $-2.0$\\
\hline
\end{tabular}
\tablefoot{The modules used in the simulations are: \texttt{[sfhdelayed]}, \texttt{[bc03]}, \texttt{[nebular]}, \texttt{[dustatt\_modified\_starburst]}, \texttt{[dl2014]}, \texttt{[skirtor2016]}, \texttt{[yang20]}, \texttt{[nebular\_AGN]}, and \texttt{[redshifting]}. Parameters not shown in the table are set to their default values.}
\label{Table:line_dust}     
\end{minipage}
\clearpage

\section{Best-fit quasar spectrum}
\label{app: best_fit_quasar_spectrum}
\noindent
\begin{minipage}{\textwidth}
\centering
\includegraphics[scale=0.5]{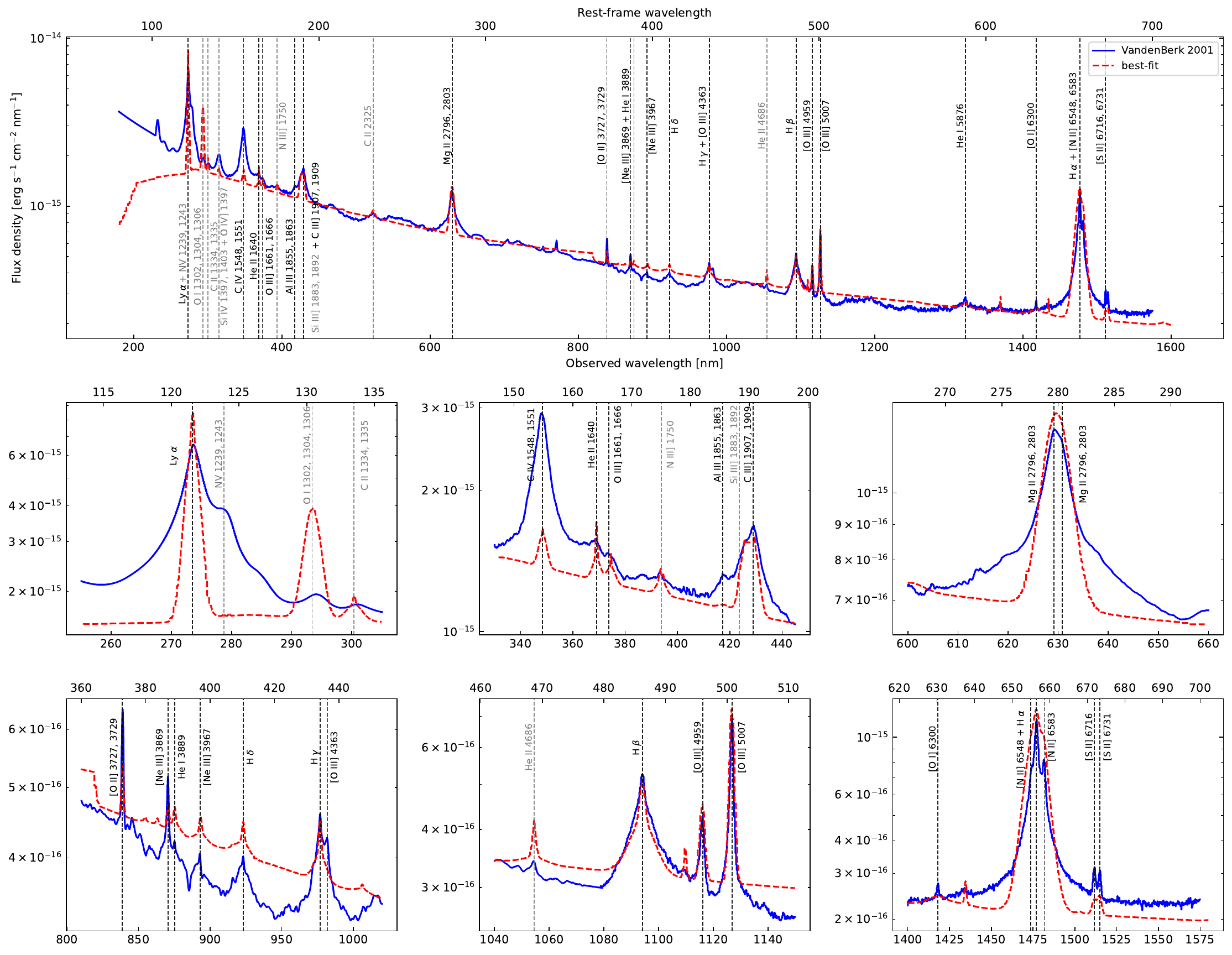}
\captionof{figure}{Comparison between the composite quasar spectrum (blue solid line) and the best-fit spectrum (red dashed line). The vertical black dashed lines indicate the emission lines used in the fitting procedure. The grey dashed lines mark emission features that appear in the composite and/or simulated spectra but are excluded from the fitting process because they are not available in the CIGALE emission-line filters. Consequently, these grey-marked emission features are neither used in the fitting nor considered in the evaluation of the fitting quality.}
\label{Fig:line profile}
\end{minipage}

\clearpage

\section{Model parameter reliability check}
\label{app: reliability}
\noindent

\begin{minipage}{\textwidth}
\centering
\includegraphics[scale=0.5]{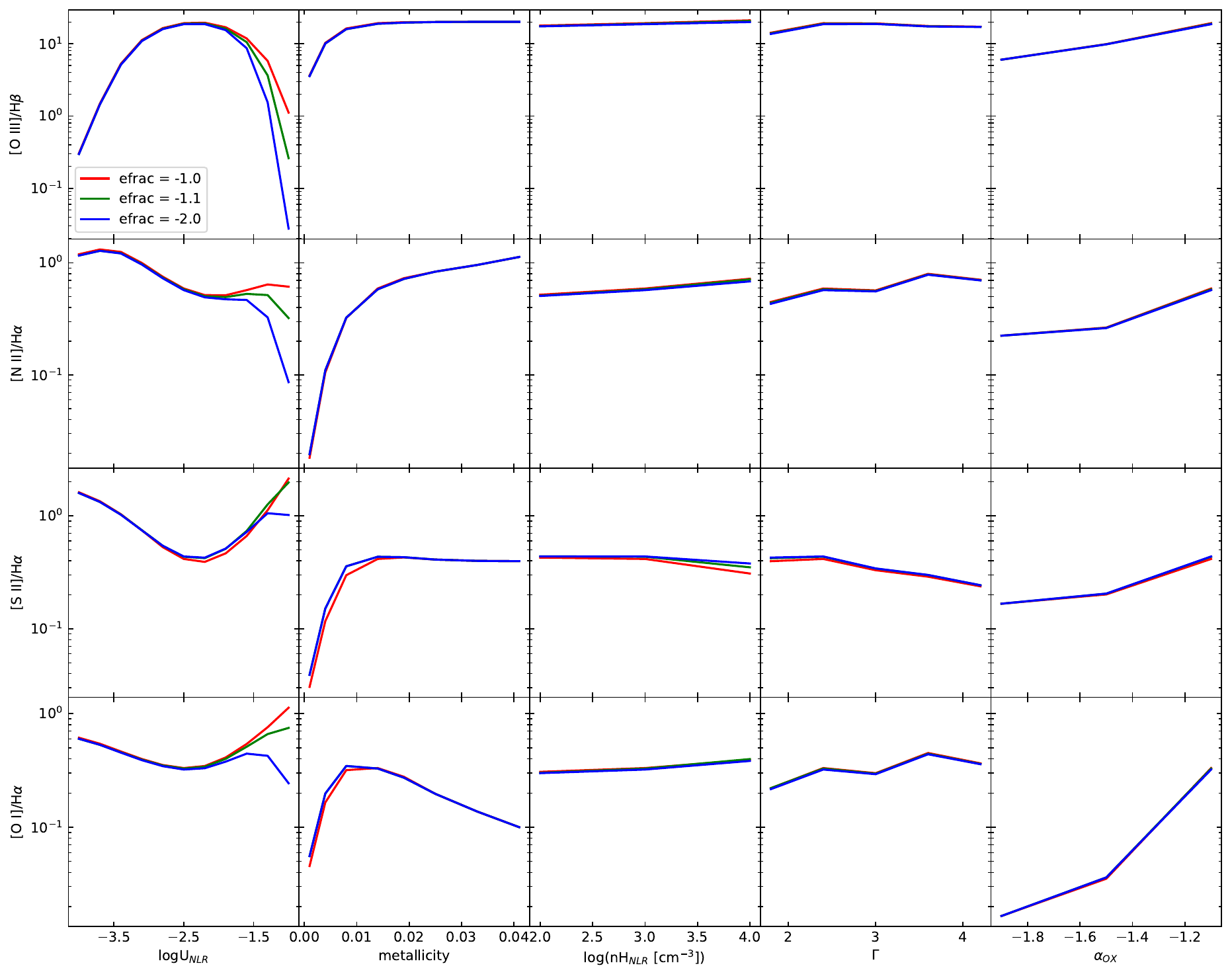}
\captionof{figure}{Sensitivity of NLR model predictions to the choice of stopping criteria.}
\label{Fig:NLR_reliability}
\end{minipage}

\begin{minipage}{\textwidth}
\centering
\includegraphics[scale=0.5]{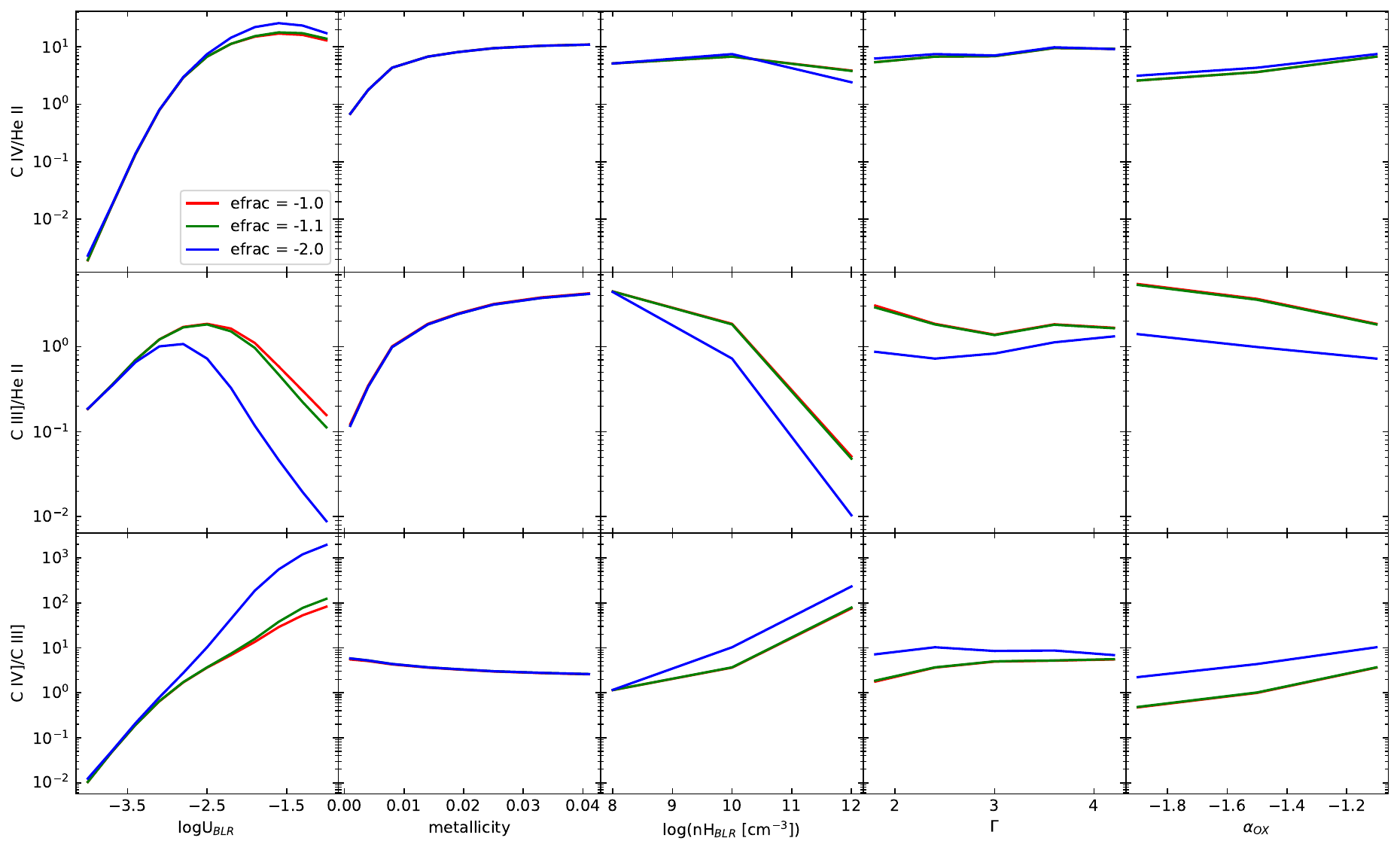}
\captionof{figure}{Sensitivity of BLR model predictions to the choice of stopping criteria.}
\label{Fig:BLR_reliability}
\end{minipage}
\clearpage

To assess the sensitivity of the parameters in the \texttt{[nebular\_AGN]} module to the choice of stopping criteria, we first selected a fiducial NLR model and a fiducial BLR model. For each parameter, we then varied it over its full range while keeping all other parameters fixed at their fiducial values and examined the line ratios predicted under three different stopping criteria, namely \texttt{efrac}=$-1.0$, $-1.1$, and $-2.0$, as a function of the varied parameter. Here, \texttt{efrac} is defined as log$(\frac{\mathrm{H^{+}}}{\mathrm{H^{0}}})$. The differences in the predicted line ratios among models adopting different stopping criteria quantify the sensitivity of the emission-line diagnostics associated with each parameter to the choice of stopping criteria.

The fiducial NLR model adopts the following parameter values: $\mathrm{\alpha_{OX}}=-1.1$, $\mathrm{\Gamma}=2.4$, $\mathrm{\delta}=0.0$, $\mathrm{logU_{NLR}}=-2.5$, $\mathrm{Z_{gas}}=0.014$, and log$\,\mathrm{nH_{NLR}}=3.0$. The reliability of the derived NLR parameters is evaluated using the four diagnostic line ratios employed in the BPT/VO87 diagrams. The fiducial BLR model adopts the following parameter values: $\mathrm{\alpha_{OX}}=-1.1$, $\mathrm{\Gamma}=2.4$, $\mathrm{\delta}=0.0$, $\mathrm{logU_{BLR}}=-2.5$, $\mathrm{Z_{gas}}=0.014$, and log$\,\mathrm{nH_{BLR}}=10.0$. The reliability of the derived BLR parameters is evaluated using three UV emission-line ratios associated with the BLR (\ion{C}{iii}]\,1907,1909/\ion{He}{ii}\,1640, \ion{C}{iv}\,1548,1551/\ion{He}{ii}\,1640, and \ion{C}{iv}\,1548,1551/\ion{C}{iii}]\,1907,1909). The results of the assessment are shown in Figs.~\ref{Fig:NLR_reliability} and~\ref{Fig:BLR_reliability}.

We find that, for the NLR models, different stopping criteria significantly affect the predicted line ratios for $\mathrm{logU_{NLR}} \geq -2.0$. For the other four parameters and for $\mathrm{logU_{NLR}} < -2.0$, the differences in the predicted line ratios among models adopting different stopping criteria are very small, generally within 0.1. This indicates that the predicted emission-line diagnostics and the corresponding parameter constraints are generally robust against the choice of stopping criteria.

For the BLR models, we find that the predictions are relatively reliable only for cases with $\mathrm{logU_{BLR}} < -3.0$ and for metallicity estimates, whereas the remaining parameters are strongly affected by the choice of stopping criteria. Although dust attenuation is not included in the BLR models, this result is still reasonable given their extremely high hydrogen densities.

It should be noted that the reliability of parameter estimates can also be affected by other factors. Since shocks, outflows, and PDRs are not included in our models, the contributions of these components to the emission lines are entirely attributed to photoionization modelled by the \texttt{[nebular\_AGN]} module, which may lead to overestimated values of some parameters. A detailed investigation of the impact of these effects is beyond the scope of this work. Moreover, although we find that the NLR models can reliably constrain $\mathrm{\Gamma}$, considering the discontinuity shown in Fig.~\ref{Fig:Feltre incident radiation field}, we recommend users restrict $\Gamma$ to values of $\leq 3.6$. Furthermore, considering the observational constraints on the photon index of the warm corona, it is preferable to restrict $\Gamma$ to values of $\leq 3.0$.

\clearpage

\section{Line sensitivity}
\label{app: line_sensitivity}
\noindent
\begin{minipage}{\textwidth}
\captionof{table}{Assessment of emission-line sensitivity to model parameters.}
\centering
\begin{tabular}{lccccccccc}
\toprule
line & fracAGN & $\alpha_{OX}$ & $\Gamma $& $\delta$ & \textbf{U}$\mathrm{_{NLR}}$ & \textbf{U}$\mathrm{_{BLR}}$ &$\mathrm{Z_{gas}}$ & $\mathrm{nH_{NLR}}$ & $\mathrm{nH_{BLR}}$\\
\midrule
\multicolumn{9}{l}{\textbf{VUV--UV}} \\
\midrule
\,Ly\,$\alpha$ & & & & & & & & & w \\
\,\ion{C}{iv}\,1548 & w & w & s & & & s & s & & s \\
\,\ion{C}{iv}\,1550 & & w & s & & & s & s & & s \\
\,\ion{He}{ii}\,1640 & w & & s &  & & & & & w \\
\,\ion{O}{iii}]\,1661* & s & w & w & & w & w & s & & s \\
\,\ion{O}{iii}]\,1666* & s & w & w & & w & w & s & & s \\
\,\ion{C}{iii}]\,1907* & w & w & w & & & & s & & w \\
\,\ion{C}{iii}]\,1909 & & w & w & & & w & s & & s \\
\,[\ion{Ne}{iv}]\,2422* & s & s & s & & s & & s & & w \\
\,[\ion{Ne}{iv}]\,2424* & s & s & s & & s & & s & & w \\
\,\ion{Mg}{ii} 2796 & & & & & & s & & & s \\
\,\ion{Mg}{ii} 2803 & & & & & & s & & & s \\
\midrule
\multicolumn{6}{l}{\textbf{Visible}} \\
\midrule
\,[\ion{Ne}{v}]\,3346* & s & s & s & & s & & s & & \\
\,[\ion{Ne}{v}]\,3426* & s & s & s & & s & w & s & & \\
\,[\ion{O}{iii}]\,3463* & s & & & & & & & & \\
\,[\ion{O}{ii}]\,3726 & & & & & s & & s & & w \\
\,[\ion{O}{ii}]\,3729 & & & & & s & & s & w & w \\
\,[\ion{Ne}{iii}]\,3869 & & & & & & & s & & s \\
\,\ion{He}{i}\,3889 & & & & & & w & & & s \\
\,[\ion{Ne}{iii}]\,3967 & & & & & & & s & & s\\
\,[\ion{O}{iii}]\,4959 & & & & & w & & w & & w \\
\,[\ion{O}{iii}]\,5007 & & & & & w & & w & & w \\
\,[\ion{O}{i}] 6300 & w & & & & & & w & & w \\
\,[\ion{O}{i}] 6364 & w & & & & & & w & & w \\
\,[\ion{N}{iii}]\,6548 & & & & & s & & s & & w \\
\,[\ion{N}{iii}]\,6583 & & & & & s & & s & & w \\
\,[\ion{S}{iii}]\,6716 & & & & & s & & & & w\\
\,[\ion{S}{iii}]\,6731 & & & & & s & & w & & w \\
\,[\ion{Ar}{iii}]\,7136 & & & & & & & & & w \\
\midrule
\multicolumn{6}{l}{\textbf{IR}} \\
\midrule
\,[\ion{S}{iii}]\,9069 & & & & & w & & s & & w \\
\,[\ion{S}{iii}]\,9532 & & & & & w & & s & & w \\
\,[\ion{Ar}{iii}]\,8.99\,$\mu$m & w & & & & & & & & w\\
\,[\ion{S}{iv}]\,10.51\,$\mu$m & w & & & & s & & s & & w \\
\,[\ion{Ne}{ii}]\,12.81\,$\mu$m & & w & & & w & w & w & & w \\
\,[\ion{Ne}{iii}]\,15.55\,$\mu$m & w & & & & & & s & & w \\
\,[\ion{S}{iii}]\,18.71\,$\mu$m & w & & & & & & s & & w \\
\,[\ion{O}{iv}]\,25.89\,$\mu$m & w & w & w & & s & & s & & w \\
\,[\ion{Fe}{ii}]\,25.98\,$\mu$m & & w & w & & w & & s & & w \\
\,[\ion{S}{iii}]\,33.47\,$\mu$m & w & & & & & & w & & w \\
\,[\ion{Si}{ii}]\,34.80\,$\mu$m & & w & & & w & & w & & w \\
\,[\ion{Ne}{iii}]\,36.00\,$\mu$m & w & & & & & & & & w \\
\,[\ion{O}{iii}]\,51.80\,$\mu$m & s & & & & w & & w & & w \\
\,[\ion{N}{iii}]\,57.32\,$\mu$m & s & & & & & & & w & w \\
\,[\ion{O}{i}]\,63.17\,$\mu$m & & s & w & & & & & & w \\
\,[\ion{O}{iii}]\,88.33\,$\mu$m & s & & & & & & w & w & w \\
\,[\ion{C}{ii}]\,158\,$\mu$m* & w & w & & & & & & & w \\
\midrule
\bottomrule
\end{tabular}
\tablefoot{`s' indicates strong sensitivity of the emission line to the parameter, while `w' indicates weak sensitivity. The emission lines marked with an asterisk are commonly used UV and optical diagnostics for identifying high-redshift AGNs. They are not initially included in the table because of their relatively low brightness, although they are sensitive to AGN parameters. Some potential AGN indicator emission lines, such as \ion{N}{iii}]\,1750, \ion{Si}{iii}]\,1888, and \ion{N}{v}\,1240, are not included in the CIGALE emission-line filters. Therefore, their sensitivity to AGN parameters cannot be assessed within this framework.}
\label{Table:tracers}   
\end{minipage}
\end{appendix}
\end{document}